\documentclass[12pt]{article}
\usepackage{booktabs}
\usepackage{url}
\usepackage[figuresright]{rotating}

\usepackage{enumerate}

\RequirePackage[authoryear]{natbib}
\RequirePackage[colorlinks,citecolor=blue,urlcolor=blue]{hyperref}

\usepackage{listings}
\usepackage{xcolor}

\usepackage[ruled,vlined]{algorithm2e}

\usepackage{setspace}

\usepackage{bbm}
\usepackage{rotating}

\usepackage{amsmath}
\usepackage{multirow}
\usepackage{latexsym,amssymb,amsmath}
\usepackage{CJK}
\usepackage{picinpar,graphicx}
\usepackage{float}
\usepackage{listings}
\usepackage{color}
\usepackage{bm}
\usepackage{mathrsfs}
\usepackage{graphicx}
\usepackage{subfigure}
\usepackage{epstopdf}

\usepackage{multirow}

\usepackage{authblk}

\usepackage{threeparttable}

\usepackage{makecell}

\usepackage{booktabs}
\usepackage{makecell}
\usepackage{tabularx}
\usepackage{array}
\usepackage{amsmath}

\newtheorem{theorem}{Theorem}
\newtheorem{lemma}{Lemma}

\newtheorem{rem}{Remark}
\newtheorem{exam}{Example}

\newcommand{\ba}{\begin{array}}
\newcommand{\ea}{\end{array}}
\newcommand{\bt}{\begin{tabular}}
\newcommand{\et}{\end{tabular}}
\newcommand{\btb}{\begin{table}}
\newcommand{\etb}{\end{table}}
\newcommand{\bc}{\begin{center}}
\newcommand{\ec}{\end{center}}
\newcommand{\bea}{\begin{eqnarray}}
\newcommand{\eea}{\end{eqnarray}}
\newcommand{\Bea}{\begin{eqnarray*}}
\newcommand{\Eea}{\end{eqnarray*}}
\newcommand{\beq}{\begin{equation}}
\newcommand{\eeq}{\end{equation}}
\newcommand{\ben}{\begin{enumerate}}
\newcommand{\een}{\end{enumerate}}

 \def \es2{$E(s^2)$}

\begin{document}

\title{A Unified Framework for Robust Blinded Sample Size Re-estimation Based on Nuisance Parameters}

\author[1]{Long-Hao Xu\footnote{Corresponding author: {e-mail: long-hao.xu@med.uni-goettingen.de}; {longhao.xu@outlook.com}}}
\author[2]{Lei Li}
\author[1]{Tim Friede}

\affil[1]{Department of Medical Statistics, University Medical Center G\"{o}ttingen, G\"{o}ttingen, Germany}
\affil[2]{LunarAI LLC, Newark, DE, USA}

\maketitle



\begin{abstract}

Blinded sample size re-estimation aims to mitigate uncertainty in nuisance parameters at the design stage while preserving the integrity of the trial. It allows the required sample size to be updated without revealing treatment assignments and thereby reduces the risk of an inappropriate sample size.
Although numerous procedures have been proposed for different outcome types, a unified framework is still lacking. We therefore develop a general framework for blinded sample size re-estimation based on nuisance parameters. Within this general framework, we develop robust methods to reduce the influence of extreme observations.
The proposed framework combines asymptotically linear estimators for the parameter of interest with robust $Z$-estimators for nuisance parameters, thereby unifying a broad class of existing procedures through different choices of estimating functions. We also propose a general principle for constructing nuisance parameters that can be estimated from blinded data. The proposed framework is supported by asymptotic theory showing that the type~I error rate converges to the nominal level under regularity conditions. Simulation studies demonstrate that the proposed method maintains reliable type~I error control across a range of settings, including contaminated data and small pilot sample sizes. Compared with existing blinded and unblinded methods, the proposed framework is more flexible and less sensitive to outliers. A clinical trial application illustrates its practical utility.

\end{abstract}


\section{Introduction}

Blinded sample size re-estimation has become a popular approach in clinical trials because it can be conducted without revealing treatment assignments, thus preserving the integrity of the trial and reducing the need for an independent data monitoring committee (e.g., see \cite{Gould:2001,Proschan:2005,FriedeKieser:2006,FriedeKieser:2013,MuetzeFriede:2021} for a detailed review).
Moreover, regulatory guidance from the European Medicines Agency and U.S. Food and Drug Administration has discussed adaptive and blinded interim procedures and has emphasized the importance of restricting access to comparative interim results in order to preserve trial integrity (e.g., \cite{ICH:1998}, \cite{EMA:2007}, \cite{FDA:2019}, \cite{ICH:2025}).
A recent systematic review (\citealp{Manoetal:2023}) identified 27 clinical trials involving sample size re-estimation, of which nine were classified as using blinded procedures, spanning different phases of clinical development and a broad range of therapeutic areas.

The idea of sample size re-estimation can be traced back to \cite{Stein:1945}. This idea was subsequently introduced into the field of clinical trials by \cite{WittesBrittain:1990}, \cite{BirkettDay:1994}, and \cite{Zuckeretal:1999}. 
A large literature has developed on blinded sample size re-estimation for a variety of endpoints, including continuous outcomes (\cite{GouldShih:1992}, \cite{KieserFriede:2003}, \cite{FriedeKieser:2006}, \cite[Chapter 11]{Proschanetal:2006}, \cite{Poschetal:2018}, \cite{MuetzeFriede:2021}), binary outcomes (\cite{Gould:1992}, \cite[Section 14.2]{JennisonTurnbull:1999}, \cite{FriedeKieser:2004}, \cite[Section 11.3]{Proschanetal:2006}, and \cite{Friedeetal:2007}), count outcomes (\cite{FriedeSchmidli:2010}, \cite{FriedeSchmidli:2010b}, \cite{Schneideretal:2013}, \cite{Schneideretal:2013b}, \cite{MuetzeFriede:2021}).

Most existing studies have primarily relied on extensive simulation experiments to assess the applicability of blinded methods, whereas comparatively little attention has been paid to the asymptotic theory for the average final sample size and type~I error control. Moreover, the limited available theoretical results have focused mainly on continuous outcomes.
\cite{Govindarajulu:2003} used total variance in the final test statistic and provided theoretical justification for type~I error control in the blinded setting. However, the results rely on restrictive assumptions, including that the mean and variance are not functionally related and that the treatment and control groups have equal variances. Such assumptions generally do not hold for binary and count outcomes, limiting the direct applicability of these results to more general outcome settings.
\cite{Proschan:2005} presented some results for continuous and binary outcomes in the blinded setting, but did not establish a general theory for type~I error control. 
For continuous outcomes, \cite{Miller:2005} investigated the properties of final variance estimation in the unblinded setting, while \cite{Poschetal:2018} examined analogous properties in the blinded setting.
In addition, \cite{GlimmLaeuter:2013} and \cite{Lu:2016} derived the exact distribution of the final $t$-test statistic under normally distributed outcomes in the blinded setting.

To the best of our knowledge, a unified theoretical framework for blinded sample size re-estimation is still lacking. Previous studies have typically considered different outcome types separately, without explicitly identifying the common rationale underlying these methods across settings. In particular, a general framework for systematically defining nuisance parameters that can be estimated from blinded data remains lacking.
Existing blinded sample size re-estimation methods generally do not explicitly account for potential outliers, even though outliers are commonly encountered in clinical trials (e.g., \cite{FarcomeniVentura:2012}).
Moreover, previous studies have largely overlooked settings involving small sample sizes, such as rare disease trials, as well as settings in which the treatment and control groups have unequal variances.

To address these gaps, we develop a unified theoretical framework for robust blinded sample size re-estimation, within which several existing methods can be recovered as special cases.
We first derive a general sample size formula based on asymptotically linear estimators of the parameters of interest. Within this framework, we further introduce a unified principle for defining nuisance parameters that can be estimated from blinded data and employ robust $Z$-estimators (e.g., \cite[Chapter 5]{VanderVaart:2000}) instead of conventional maximum likelihood estimators to estimate these parameters from the blinded internal pilot data.
We further demonstrate that the lumping and mixture approaches, both commonly used in blinded sample size re-estimation (see Remark \ref{remark-1}), can be formulated as special cases of the proposed $Z$-estimation framework. To illustrate its generality and applicability across different outcome types, we present several examples showing how existing procedures can be recovered and new robust sample size re-estimation methods can be developed within the same framework.
From a theoretical perspective, we establish the asymptotic properties of the proposed procedure. In particular, we demonstrate that the final sample size is asymptotically equivalent to the target sample size induced by the chosen blinded estimating equation and that the nominal type~I error level is asymptotically preserved under the null hypothesis, subject to suitable regularity conditions. Simulation studies show that the proposed method generally maintains type~I error rates close to the nominal level across a range of scenarios. In the presence of contamination, robust estimation can reduce the influence of extreme observations, thereby improving the stability of the final sample size.

The remainder of the paper is organized as follows.
Section \ref{sec-2} introduces a unified theoretical framework for robust blinded sample size re-estimation and establishes the corresponding asymptotic results.
Section \ref{sec-3} presents extensive simulation studies to assess the finite-sample performance of the proposed method.
Section \ref{sec-4} provides a clinical trial application to demonstrate its practical utility.
Section \ref{sec-5} concludes with a discussion. Technical proofs are deferred to Section A of Supplementary Materials.

\section{Robust blinded sample size re-estimation}\label{sec-2}

In this section, we propose a new unified framework for robust blinded sample size re-estimation and establish its asymptotic properties. In Section B of Supplementary Materials, we present several examples demonstrating that existing methods in the literature arise as special cases of the proposed framework. In addition, we derive new robust sample size re-estimation procedures within this unified framework.

\subsection{Proposed general framework}

Consider a two-sample problem comparing parameters from two populations.
Assume that $X_1,X_2,\ldots$ from the treatment group are independent and identically distributed (i.i.d.) with distribution $F(\theta_1,\omega)$, and, independently, $Y_1,Y_2,\ldots$ from the control group are i.i.d. with distribution $F(\theta_2,\omega)$.
We assume a $r:1$, $r\in\mathbb{N}^+$ allocation ratio between the treatment group and the control group, which is common in clinical trials.
Here $\theta_1,\theta_2\in\mathbb R$ denote the parameters of primary interest and $\omega\in\mathbb R^q$ denotes the shared parameters in the two groups.
We consider the null hypothesis
\begin{align}\label{h0}
H_0:\ \theta_1-\theta_2=0.
\end{align}

We further assume that asymptotically linear estimators are available.
Specifically, under $F(\theta_1,\omega)$, suppose there exists an estimator $\widehat{\theta}_{1,m}=\widehat{\theta}_{1,m}(X_1,\ldots,X_m)$ such that
\begin{align}\label{asymptotically-linear-P}
\sqrt{m}(\widehat{\theta}_{1,m}-\theta_1)=\frac{1}{\sqrt{m}}\sum_{i=1}^m \varphi_1(X_i;\theta_1,\omega) + o_p(1),
\end{align}
where $\mathbb{E}(\varphi_1(X_i))=0$ and $0<\mathbb{E}(\varphi_1^2(X_i;\theta_1,\omega))\equiv \tau^2_1 <\infty$.
Analogously, we assume that, based on the $Y_j$ under $F(\theta_2,\omega)$,
\begin{align}\label{asymptotically-linear-Q}
\sqrt{n}(\widehat{\theta}_{2,n}-\theta_2)=\frac{1}{\sqrt{n}}\sum_{j=1}^n \varphi_2(Y_j;\theta_2,\omega) + o_p(1),
\end{align}
where $\mathbb{E}(\varphi_2(Y_j))=0$ and $0<\mathbb{E}(\varphi_2^2(Y_j;\theta_2,\omega))\equiv \tau^2_2 <\infty$.
For example, commonly used estimators, including the sample mean, sample variance, and sample quantiles, admit asymptotically linear representations under suitable regularity conditions (e.g., \cite{Serfling:1980}).

From the central limit theorem, it follows
\begin{align*}
\sqrt{m}(\widehat{\theta}_{1,m}-\theta_1)  \overset{d}{\rightarrow} \mathcal{N}(0, \tau^2_1), \ \mbox{and} \
\sqrt{n}(\widehat{\theta}_{2,n}-\theta_2)  \overset{d}{\rightarrow} \mathcal{N}(0, \tau^2_2).
\end{align*}
Note that $\tau_1^2$ is the asymptotic variance of $\sqrt{m}(\widehat\theta_{1,m}-\theta_1)$, and $\tau_2^2$ is the asymptotic variance of $\sqrt{n}(\widehat\theta_{2,n}-\theta_2)$.

Consider testing the null hypothesis (\ref{h0}) based on a test statistic of the form
\begin{align*}
T_{m,n} = \frac{(m+n)^{1/2}(\widehat{\theta}_{1,m}(X_1,\ldots,X_m)-\widehat{\theta}_{2,n}(Y_1,\ldots,Y_n))}{\sqrt{(m+n)(\tau_1^2/m+\tau_2^2/n)}},
\end{align*}
where the estimators $\widehat{\theta}_{1,m}(X_1,\ldots,X_m)$ and $\widehat{\theta}_{2,n}(Y_1,\ldots,Y_n)$ satisfy (\ref{asymptotically-linear-P}) and (\ref{asymptotically-linear-Q}), respectively.
Under the null hypothesis $H_0$ in (\ref{h0}), we have $T_{m,n} \overset{d}{\rightarrow} \mathcal{N}(0,1)$.
Under the alternative hypothesis $H_1: \theta_1-\theta_2=\delta_1$, the statistic $T_{m,n}$ approximately follows a normal distribution $\mathcal{N}({\delta_1\sqrt{m+n}}/{\sqrt{(m+n)(\tau_1^2/m+\tau_2^2/n)}},1)$.

To plan a fixed design achieving a one-sided type I error rate $\alpha$ and power $1-\beta$ to detect $\delta_1=\theta_1-\theta_2$ under the alternative, if $\tau_1^2$ and $\tau_2^2$ are known, we require approximately
\begin{align*}
1-\Phi\left(z_{1-\alpha}-\frac{|\delta_1|\sqrt{m+n}}{\sqrt{(m+n)(\tau_1^2/m+\tau_2^2/n)}}\right) \ge 1-\beta,
\end{align*}
where $\Phi(\cdot)$ denotes the cumulative distribution function of the standard normal distribution and $z_{1-\alpha}$ is the $(1-\alpha)$ quantile of the standard normal distribution.
Solving for $n$ with $m=rn$ yields the usual sample size formula for the control group
\begin{align}\label{n}
n=\frac{(z_{1-\alpha}+z_{1-\beta})^2}{\delta^2_1}\left(\frac{\tau_1^2}{r}+\tau_2^2\right),
\end{align}
where $z_{1-\beta}$ is the $(1-\beta)$ quantile of the standard normal distribution.
In practice, $\delta_1$ is typically chosen as the minimal clinically important difference.

\begin{rem}(General sample size formula)

Equation (\ref{n}) provides a unified formula that can be applied to a variety of outcome types, including continuous, binary, and count outcomes.
When the true values of $\tau_1^2$ and $\tau_2^2$ are known, Equation (\ref{n}) gives the required sample size for a fixed-sample design.
In practice, the true variances $\tau_1^2$ and $\tau_2^2$ are unknown.
\end{rem}

Following the conventional framework for blinded sample size re-estimation (e.g., \cite{FriedeKieser:2013}), we introduce the following basic notation. 
Let $m_1$ and $n_1$ denote the numbers of patients in the treatment and control groups of the internal pilot study, respectively, yielding observations $X_1,\ldots,X_{m_1}$ and $Y_1,\ldots,Y_{n_1}$. Denote the total internal pilot sample size by $n_p = m_1 + n_1$. We further assume a fixed allocation ratio $m_1 / n_1 = r \in \mathbb{N}^+$. Accordingly, the total internal pilot sample size is $n_p = m_1 + n_1 = (1+r)n_1$.
These pilot data are then used to obtain updated estimates of $\tau_1^2$ and $\tau_2^2$, which are plugged into the sample size formula to determine the planned final sample size.

Define the blinded internal pilot data vector as
\begin{align}\label{blinded-vec}
\mathbf{Z}=(Z_1,\ldots,Z_{n_p}),
\end{align}
where the treatment assignment labels are concealed.
Of the $n_p$ observations, $m_1$ originate from the treatment group and $n_1$ from the control group, but their group memberships remain unknown under blinding.

To construct nuisance parameters that remain identifiable under blinding, we propose a reparameterization principle based on a prespecified effect measure and a blinded summary statistic.
Let $\delta_1$ be the target effect size used for planning. 
We formulate the following system of equations:
\begin{equation}
\label{eq:reparam_system_general}
\begin{cases}
\theta_1-\theta_2=\delta_1,\\
\mathbb{E}_{W_1\sim F(\bar\theta,\omega)}(W_1)=\mathbb{E}_{W_2\sim \frac{r}{r+1}F(\theta_1,\omega)+ \frac{1}{r+1}F(\theta_2,\omega)}(W_2).
\end{cases}
\end{equation}
The first equation reflects the design requirement of achieving power $1-\beta$ at the prespecified effect size $\delta_1$, where $\delta_1$ represents the minimum clinically important difference.
The second equation follows from the moment-matching condition.

To ensure that the group-specific parameters can be uniquely determined from the nuisance parameters estimable from blinded data and the prespecified treatment effect, we impose the following identifiability assumption.
For each admissible $(\bar\theta,\omega,\delta_1)$ in the parameter space,
the system \eqref{eq:reparam_system_general} admits a unique solution
$(\theta_1,\theta_2)$ satisfying $(\theta_1,\theta_2)\in\Theta_1\times\Theta_2$.
Under this assumption, the group-specific parameters can be represented as
functions of $(\bar\theta,\omega,\delta_1)$, namely
\begin{align}\label{eq-g1g2}
\theta_1=g_1(\bar\theta,\omega,\delta_1),\qquad
\theta_2=g_2(\bar\theta,\omega,\delta_1).
\end{align}

Therefore, under the blinded setting, the estimated required sample size formula for the control group can be expressed as
\begin{align}\label{n-eta}
n(\widehat{\bm{\eta}})=\frac{(z_{1-\alpha}+z_{1-\beta})^2}{\delta^2_1}\left(\frac{\tau_1^2(\widehat{\bm{\eta}})}{r}+\tau_2^2(\widehat{\bm{\eta}})\right),
\end{align}
where $\widehat{\bm{\eta}}=(\widehat{\bar{\theta}},\widehat{\omega})$ denotes the estimator of $\bm{\eta}$ obtained from the internal pilot study.

The nuisance parameter $\bm{\eta}$ is estimated from the blinded internal pilot observations $Z_1,\ldots,Z_{n_p}$, by using a $Z$-estimator (e.g., \cite[Chapter 5]{VanderVaart:2000}).
Let $\Theta\subseteq\mathbb{R}^{q+1}$ denote the parameter space and $\mathcal{Z}\subseteq\mathbb{R}$ the observation space.
For a prespecified measurable estimating function $\Psi:\Theta\times\mathcal{Z}\to\mathbb{R}^{q+1}$, the $Z$-estimator $\widehat{\bm{\eta}}\in\Theta$ is defined as any measurable solution to the vector-valued estimating equation
\begin{align}\label{z-estimator}
\frac{1}{n_p}\sum_{k=1}^{n_p}
\Psi(\widehat{\bm{\eta}};Z_k)
=\mathbf{0}_{q+1},
\end{align}
where $\mathbf{0}_{q+1}$ denotes the $(q+1)$-dimensional zero vector.

\begin{rem}(A unified interpretation of the lumping and mixture approaches)\label{remark-1}

Under suitable regularity conditions, $M$-estimators can be expressed as special cases of $Z$-estimators through their first-order optimality conditions. Specifically, an $M$-estimator is defined as a minimizer of the empirical loss function,
\begin{align*}
\widehat{\bm{\eta}}
\in \arg\min_{\bm{\eta}\in\Theta}
\frac{1}{n_p}\sum_{i=1}^{n_p}
\rho(Z_i,\bm{\eta}),
\end{align*}
where $\rho:\mathcal{Z}\times\Theta\to\mathbb{R}$ is a prespecified loss function.
If $\rho$ is differentiable with respect to $\bm{\eta}$ and the minimizer lies in the interior of $\Theta$, the corresponding estimating function is given by $\Psi(\bm{\eta};z)=\nabla_{\bm{\eta}}\rho(z,\bm{\eta})$.
In particular, choosing $\rho(z,\bm{\eta})=-\log f(z;\bm{\eta})$, where $f$ denotes an assumed probability density or mass function, yields the maximum likelihood estimator.

In the literature on blinded sample size re-estimation, two main approaches have been used to specify the working density $f$ for nuisance parameter estimation from blinded internal pilot data: the lumping approach (e.g., \cite{FriedeSchmidli:2010}, \cite{FriedeSchmidli:2010b}, \cite{Schneideretal:2013}, \cite{Friedeetal:2019}, \cite{MuetzeFriede:2021}, \cite[Chapter 11]{Proschanetal:2006}, \cite{KieserFriede:2003}, \cite{Zuckeretal:1999}) and the mixture approach (e.g., \cite{Asendorfetal:2019}, \cite{Asendorfetal:2019b}, \cite{Muetzeetal:2020}, \cite{Zapfetal:2020}).

In the lumping approach, the blinded internal pilot observations are modeled using a single working distribution $F(\bar{\theta}_L,\omega_L)$, with the nuisance parameters estimated by maximizing the corresponding likelihood. The associated loss function is defined as
\begin{align*}
\rho(z_i,\bm{\eta}_L)
=
-\log f(z_i;\bar{\theta}_L,\omega_L),
\end{align*}
where $\bm{\eta}_L=(\bar{\theta}_L,\omega_L)$ denotes the nuisance parameter vector, and $f(\cdot;\bar{\theta}_L,\omega_L)$ is the probability density or mass function associated with the working distribution $F(\bar{\theta}_L,\omega_L)$.

In the mixture approach, the blinded internal pilot observations are modeled using a two-component working mixture distribution of the form
\[
\frac{r}{r+1}F(\theta_1',\omega_M)
+
\frac{1}{r+1}F(\theta_2',\omega_M),
\]
where
$\theta_1'=g_{1,M}(\bar{\theta}_M,\omega_M,\delta_1)$ and
$\theta_2'=g_{2,M}(\bar{\theta}_M,\omega_M,\delta_1)$
are determined by the reparameterization in \eqref{eq-g1g2}.
The nuisance parameters are estimated by maximizing the corresponding working likelihood. The associated loss function is defined as
\begin{align*}
\rho(z_i,\bm{\eta}_M)
=
-\log\left(
\frac{r}{r+1}f(z_i;\theta_1',\omega_M)
+
\frac{1}{r+1}f(z_i;\theta_2',\omega_M)
\right),
\end{align*}
where $\bm{\eta}_M=(\bar{\theta}_M,\omega_M)$ denotes the nuisance parameter vector, and $f(\cdot;\theta_j',\omega_M)$ denotes the probability density or mass function associated with $F(\theta_j',\omega_M)$, for $j=1,2$.

Within the proposed framework, both the lumping and mixture approaches can be formulated as special cases of $M$-estimation, differing only in the specification of their respective loss functions $\rho$.

\end{rem}

Using the nuisance parameter estimator $\widehat{\bm{\eta}}$ obtained from the blinded internal pilot data through \eqref{z-estimator}, we define the final sample size for the control group as
\begin{align}\label{eq-N}
N = \max\left\{
\lfloor n(\widehat{\bm{\eta}})\rfloor+1,\; n_1
\right\},
\end{align}
where $\lfloor u\rfloor$ denotes the largest integer not exceeding $u$. The corresponding final sample size for the treatment group is $M=rN$, where $r\in\mathbb{N}^+$ is the prespecified allocation ratio.
If $N>n_1$, an additional $M-m_1$ and $N-n_1$ patients are recruited to the treatment and control groups, respectively, yielding additional observations $X_{m_1+1},\ldots,X_M$ and $Y_{n_1+1},\ldots,Y_N$. Otherwise, $N=n_1$, and no further recruitment is required.

Upon completion of the trial, treatment assignments are revealed, and the final analysis is performed using all available observations. We consider the Wald-type test statistic
\begin{align}\label{test-statistics}
T_{M,N}
=
\frac{
\sqrt{M+N}
\left(
\widehat{\theta}_{1,M}(X_1,\ldots,X_M)
-
\widehat{\theta}_{2,N}(Y_1,\ldots,Y_N)
\right)
}{
\sqrt{
(M+N)
\left(
\widehat{\tau}_{1,0}^2/M
+
\widehat{\tau}_{2,0}^2/N
\right)
}
},
\end{align}
where $\widehat{\tau}_{1,0}^2$ and $\widehat{\tau}_{2,0}^2$ are consistent estimators of $\tau_1^2$ and $\tau_2^2$, respectively, based on the final unblinded data.
The rejection region is determined by the direction of the prespecified alternative hypothesis.
For a positive target effect $\delta_1>0$, corresponding to $H_1:\theta_1>\theta_2$, the null hypothesis is rejected if $T_{M,N}>z_{1-\alpha}$.
Conversely, for a negative target effect $\delta_1<0$, corresponding to $H_1:\theta_1<\theta_2$, the null hypothesis is rejected if $T_{M,N}<z_{\alpha}$.

In summary, the proposed robust blinded sample size re-estimation procedure comprises two main components: monitoring and testing. The monitoring component estimates nuisance parameters from the blinded internal pilot data and uses these estimates to update the final sample size. The testing component performs statistical inference using the complete trial data after unblinding. The overall procedure is outlined in Algorithm \ref{alg:robust-bsr}.

\begin{algorithm}[H]
\caption{Robust Blinded Sample Size Re-estimation Procedure}
\label{alg:robust-bsr}

\KwIn{Blinded internal pilot data, prespecified allocation ratio $r$, and internal pilot sample sizes $n_1$ (control) and $m_1=rn_1$ (treatment)}
\KwOut{Final sample sizes $N$ and $M$, and the final test decision}

\BlankLine
\textbf{Step 1: Internal Pilot Study.}
Collect the blinded internal pilot observations $\mathbf{Z}=(Z_1,\ldots,Z_{n_p})$ from the ongoing trial.

\BlankLine
\textbf{Step 2: Robust Nuisance Parameter Estimation.}
Obtain the robust nuisance parameter estimator $\widehat{\bm{\eta}}$ by solving the estimating equation \eqref{z-estimator} using the blinded pilot observations $\mathbf{Z}$.

\BlankLine
\textbf{Step 3: Sample Size Re-estimation.}
Compute the updated sample size target $n(\widehat{\bm{\eta}})$ using \eqref{n-eta}.
Determine the final sample sizes $N$ and $M=rN$ according to \eqref{eq-N}.
If $N>n_1$, recruit an additional $N-n_1$ patients to the control group and $M-m_1$ patients to the treatment group.

\BlankLine
\textbf{Step 4: Final Analysis.}
Upon completion of recruitment, unblind the trial data and compute the Wald-type test statistic $T_{M,N}$ defined in \eqref{test-statistics} using the complete dataset $\{X_1,\ldots,X_M,Y_1,\ldots,Y_N\}$.
Make the final test decision according to the prespecified rejection rule.

\end{algorithm}

\begin{rem}(Application to different blinded sample size re-estimation procedures)

To demonstrate the generality of the proposed framework, we show in Section B of Supplementary Materials how several existing blinded sample size re-estimation methods can be recovered as special cases and derive new robust procedures within the same framework. For each example, we specify the nuisance parameters that can be estimated from blinded data and solve the reparameterization system~(\ref{eq:reparam_system_general}) to obtain the group-specific parameters required for sample size planning.
\end{rem}

\subsection{Theoretical results}

Under the regularity conditions given in Section A of Supplementary Materials, we establish the asymptotic properties of the proposed method.
Let $\bm{\eta}\in\mathbb{R}^{q+1}$ denote the nuisance parameter, and let $\widehat{\bm{\eta}}$ be its estimator based on the blinded internal pilot data of size $n_p=m_1+n_1=(1+r)n_1$, obtained from the $Z$-estimator defined in (\ref{z-estimator}).
The proofs are presented in Section A of Supplementary Materials.

\begin{lemma}(Asymptotic properties of the final sample size)\label{lemma-1}
Under Assumptions N1--N5 and R1--R4 in Section A of Supplementary Materials, we have
\begin{align*}
\frac{N}{n(\bm{\eta}_0)}\xrightarrow{p}1 \ \mbox{and} \ 
\mathbb{E}\left(\frac{N}{n(\bm{\eta}_0)}\right) \rightarrow 1, \quad \mbox{as} \ n(\bm{\eta}_0)\rightarrow\infty.
\end{align*}
\end{lemma}

Note that $\bm{\eta}_0$ is the unique solution to the weighted population estimating equation
\begin{align*}
\Psi_0(\bm{\eta})
=
\frac{r}{r+1}
\mathbb{E}_{\theta_1,\omega}(\Psi(\bm{\eta};X))
+
\frac{1}{r+1}
\mathbb{E}_{\theta_2,\omega}(\Psi(\bm{\eta};Y))
=0.
\end{align*}
The sample size $n(\bm{\eta}_0)$ should therefore be interpreted as the asymptotic target sample size induced by the chosen blinded estimating equation.

\begin{lemma}(Sharp distribution-free bound)
\label{lemma-2}
Assume that $n(\widehat{\bm{\eta}})>0$ almost surely and $\mathbb{E}(n(\widehat{\bm{\eta}})^2)<\infty$. 
Then, we have
\begin{align*}
    \sqrt{\operatorname{Var}(N)}
    \le
    \sqrt{\operatorname{Var}(n(\widehat{\bm{\eta}}))}
    +\frac{1}{2},
\end{align*}
and the constant $\frac{1}{2}$ is optimal.
\end{lemma}

Lemma \ref{lemma-2} provides a sharp finite-sample bound and shows that implementing the continuous target sample size $n(\widehat{\bm{\eta}})$ as the integer-valued final sample size $N$ does not introduce uncontrolled additional variability. Note that this result does not require a specific distributional assumption or an asymptotic approximation.

Under the null hypothesis (\ref{h0}), we will show that the test statistic $T_{M,N}$ converges in distribution to a standard normal distribution. 

\begin{theorem}(Asymptotic normality)\label{theorem-1}
Assume N1--N5 and R1--R4 in Section A of Supplementary Materials. 
Let
\begin{align*}
N=\max\{\lfloor n(\widehat{\bm\eta})\rfloor+1,\; n_1\},
\qquad
M= rN,
\end{align*}
for some fixed $r\in\mathbb{N}^+$.
Under the null hypothesis $H_0:\theta_1=\theta_2$, we have
\begin{align*}
\sqrt{M+N}\,
(\widehat\theta_{1,M}-\widehat\theta_{2,N})
\overset{d}{\rightarrow}
\mathcal{N}\!\left(
0,
\left(1+\frac1r\right)\tau_1^2+(1+r)\tau_2^2
\right).
\end{align*}
Furthermore, if the final variance estimators satisfy $\widehat\tau_{1,0}^2 \overset{p}{\rightarrow} \tau_1^2$ and $\widehat\tau_{2,0}^2 \overset{p}{\rightarrow} \tau_2^2$, then $T_{M,N}\xrightarrow{d}\mathcal{N}(0,1)$ under $H_0$.

\end{theorem}

For the one-sided alternative $H_1:\theta_1-\theta_2>0$, define the test function
\begin{align*}
\phi_{M,N}
=
\mathbb{I}\{T_{M,N}>z_{1-\alpha}\},
\end{align*}
where $z_{1-\alpha}$ is the $(1-\alpha)$-quantile of the standard normal distribution. For any fixed parameter configuration satisfying the null hypothesis $H_0:\theta_1-\theta_2=0$ and the assumptions of Theorem~\ref{theorem-1}, the convergence in distribution established therein, together with the continuity of the standard normal distribution function $\Phi$, implies
\begin{align*}
\mathbb{E}_{H_0}(\phi_{M,N})
=
\mathbb{P}_{H_0}(T_{M,N}>z_{1-\alpha})
\rightarrow
1-\Phi(z_{1-\alpha})
=
\alpha.
\end{align*}
Thus, the proposed test asymptotically preserves the nominal type I error level.

\section{Simulation}\label{sec-3}

In this section, we evaluate the performance of the proposed methods for both continuous and count outcomes.
In particular, we investigate their robustness under contamination by adopting Huber's $\varepsilon$-contamination model (cf. \cite{Huber:1964} and \cite{Huber:1965}).
In the context of a randomized controlled trial, this model reflects the presence of participants whose outcome distribution differs from that of the majority, even when randomization is valid. 
In general, Huber's $\varepsilon$-contamination model is defined as
\begin{align*}
P_\varepsilon = (1-\varepsilon)P + \varepsilon H, \qquad 0\le \varepsilon <1,
\end{align*}
where $P$ denotes the target distribution, $H$ represents the contaminating distribution that generates outliers, and $\varepsilon$ represents the proportion of contamination.

In the simulations, we consider a treatment-arm-only contamination mechanism, which is clinically interpretable as a treatment-specific deviation.
Throughout the simulations, we assume a 1:1 allocation ratio between the two groups. To examine performance across different degrees of contamination, we consider $\varepsilon = 0, 0.1, 0.2,$ and $0.3$, where $\varepsilon = 0$ corresponds to the uncontaminated setting. We also vary the pilot-study sample size by setting $n_1 = 5, 10,$ and $50$.
We compare the blinded one-sample estimator (denoted OS, cf. \cite{FriedeKieser:2013} and \cite{MuetzeFriede:2021}), the unblinded estimator (denoted UN, cf. \cite{FriedeKieser:2013} and \cite{MuetzeFriede:2021}), and the blinded robust estimator based on Huber’s Proposal 2 (denoted Robust).
Note that Huber’s Proposal 2 is naturally formulated as a $Z$-estimator (cf. \cite{Huber:1964} and \cite{HuberRonchetti:2009}). 
For Huber’s Proposal 2, we examine the impact of the tuning parameter $k$; a brief discussion is provided in Section~\ref{sec-5}.
For each scenario, we generate 100{,}000 simulated datasets. The nominal significance level is set to $\alpha = 0.025$, and the target type~II error rate is set to $\beta = 0.2$. 
The Monte Carlo standard error for an empirical type I error rate around 0.025 is approximately $4.9\times 10^{-4}$ based on 100,000 simulations.
For each method, we report the mean and standard deviation of the final sample size $N$ in the control group, together with the empirical type~I error rate and empirical power.
In practice, we use the $t$-distribution critical value $t_{1-\alpha,M+N-2}$ instead of the standard normal critical value $z_{1-\alpha}$ for the final hypothesis test in the case of small sample sizes (e.g., \cite{FriedeKieser:2006}).
Sensitivity analyses for the unequal-variance and unequal-dispersion scenarios are provided in Section D of Supplementary Materials.

To compare two methods when both power and average sample size differ, motivated by \cite{JennisonTurnbull:2006}, we consider a relative efficiency measure that jointly reflects statistical performance and sampling cost. Let $p_A$ and $p_B$ denote the powers of Methods A and B, respectively, and let $ASN_A$ and $ASN_B$ denote their corresponding average sample sizes. We define the relative efficiency ratio of Method A to Method B as
\begin{align*}
ER_{A,B}
=
\frac{ASN_B}{ASN_A}
\cdot
\frac{\left(z_\alpha^*+\Phi^{-1}(p_A)\right)^2}
{\left(z_\alpha^*+\Phi^{-1}(p_B)\right)^2}
\times 100,
\end{align*}
where $\Phi^{-1}(\cdot)$ denotes the inverse cumulative distribution function of the standard normal distribution, and $z_\alpha^*$ is the critical value corresponding to the significance level $\alpha$ (i.e., $z_{1-\alpha}$ for a one-sided test and $z_{1-\alpha/2}$ for a two-sided test). This criterion adjusts the comparison in average sample size for differences in attained power.
Values greater than 100 indicate a more favorable sample-size-to-power trade-off under this summary measure.
Additional simulation results are provided in Section D of Supplementary Materials.

\subsection{Results for continuous outcome}

As described in the previous section, the sample size for the control group is determined so that the corresponding test attains a one-sided type I error rate of $\alpha$ under the null hypothesis $H_0:\mu_1-\mu_2=0$, and power $1-\beta$ to detect an alternative corresponding to $\delta_1=0.5$.

We assume that the true data-generating mechanism is given by
\begin{align*}
X_i \sim (1-\varepsilon)\mathcal{N}(\mu_1,\sigma_1^2) + \varepsilon \mathcal{N}(\mu_1,4^2), \qquad
Y_j \sim \mathcal{N}(\mu_2,\sigma_2^2),
\end{align*}
where $\mathcal{N}(\mu_1,4^2)$ represents the contaminating distribution that generates outliers.

\begin{table}[htbp]
\centering
\footnotesize
\caption{Simulation under the null hypothesis ($\mu_1=\mu_2=0$, $\sigma_1=\sigma_2=1$)}
\label{tab:combined_type1_sample_h0_sigma11}
\resizebox{\textwidth}{!}{
\begin{tabular}{llccccccccc}
\toprule
& & \multicolumn{3}{c}{$n_1=5$}
& \multicolumn{3}{c}{$n_1=10$}
& \multicolumn{3}{c}{$n_1=50$} \\
\cmidrule(lr){3-5} \cmidrule(lr){6-8} \cmidrule(lr){9-11}
$\epsilon$ & Method
& Type I & Mean($N$) & SD($N$)
& Type I & Mean($N$) & SD($N$)
& Type I & Mean($N$) & SD($N$) \\
\midrule

\multirow{6}{*}{$0$}
& OS
& 0.02447 & 63.14 & 29.57
& 0.02587 & 63.35 & 20.48
& 0.02486 & 63.47 & 8.61 \\

& UN
& 0.02678 & 63.18 & 31.38
& 0.02712 & 63.33 & 21.04
& 0.02691 & 63.47 & 8.64 \\

& Robust ($k=1.5$)
& 0.02482 & 69.86 & 36.25
& 0.02503 & 66.72 & 24.59
& 0.02502 & 64.26 & 10.00 \\

& Robust ($k=1.345$)
& 0.02555 & 71.55 & 38.60
& 0.02545 & 67.50 & 25.99
& 0.02521 & 64.48 & 10.47 \\

& Robust ($k=1.2$)
& 0.02415 & 73.46 & 41.56
& 0.02479 & 68.44 & 27.72
& 0.02486 & 64.75 & 11.02 \\

\midrule

\multirow{6}{*}{$0.1$}
& OS
& 0.02452 & 110.43 & 120.57
& 0.02462 & 110.47 & 86.74
& 0.02422 & 110.18 & 38.44 \\

& UN
& 0.02456 & 110.45 & 121.94
& 0.02590 & 110.52 & 87.36
& 0.02539 & 110.18 & 38.49 \\

& Robust ($k=1.5$)
& 0.02349 & 83.43 & 52.56
& 0.02453 & 76.14 & 30.51
& 0.02452 & 71.74 & 12.24 \\

& Robust ($k=1.345$)
& 0.02332 & 83.67 & 51.42
& 0.02431 & 76.14 & 31.18
& 0.02430 & 71.37 & 12.55 \\

& Robust ($k=1.2$)
& 0.02350 & 84.57 & 52.41
& 0.02473 & 76.52 & 32.45
& 0.02453 & 71.16 & 13.02 \\

\midrule

\multirow{6}{*}{$0.2$}
& OS
& 0.02511 & 155.91 & 161.98
& 0.02422 & 156.92 & 118.35
& 0.02448 & 157.35 & 52.81 \\

& UN
& 0.02539 & 155.72 & 163.92
& 0.02489 & 156.90 & 119.29
& 0.02607 & 157.35 & 52.89 \\

& Robust ($k=1.5$)
& 0.02495 & 101.17 & 76.59
& 0.02413 & 88.21 & 39.15
& 0.02436 & 81.11 & 14.78 \\

& Robust ($k=1.345$)
& 0.02491 & 99.12 & 70.86
& 0.02440 & 87.01 & 38.34
& 0.02429 & 79.88 & 14.82 \\

& Robust ($k=1.2$)
& 0.02510 & 98.39 & 68.68
& 0.02432 & 86.54 & 38.71
& 0.02478 & 79.02 & 15.13 \\

\midrule

\multirow{6}{*}{$0.3$}
& OS
& 0.02495 & 202.83 & 193.40
& 0.02497 & 205.00 & 141.63
& 0.02626 & 204.83 & 63.62 \\

& UN
& 0.02506 & 202.29 & 195.82
& 0.02551 & 205.00 & 143.11
& 0.02562 & 204.83 & 63.74 \\

& Robust ($k=1.5$)
& 0.02561 & 124.95 & 106.11
& 0.02450 & 103.38 & 50.58
& 0.02464 & 92.54 & 17.95 \\

& Robust ($k=1.345$)
& 0.02551 & 119.60 & 95.95
& 0.02446 & 100.20 & 47.40
& 0.02474 & 90.11 & 17.56 \\

& Robust ($k=1.2$)
& 0.02517 & 116.39 & 90.50
& 0.02501 & 98.38 & 46.33
& 0.02480 & 88.33 & 17.60 \\

\bottomrule
\end{tabular}
}
\end{table}

\begin{table}[htbp]
\centering
\footnotesize
\caption{Simulation under the alternative hypothesis ($\mu_1=0.5$, $\mu_2=0$, $\sigma_1=\sigma_2=1$)}
\label{tab:combined_power_sample_h1_sigma11}
\resizebox{\textwidth}{!}{
\begin{tabular}{llccccccccc}
\toprule
& & \multicolumn{3}{c}{$n_1=5$}
& \multicolumn{3}{c}{$n_1=10$}
& \multicolumn{3}{c}{$n_1=50$} \\
\cmidrule(lr){3-5} \cmidrule(lr){6-8} \cmidrule(lr){9-11}
$\epsilon$ & Method
& Power & Mean($N$) & SD($N$)
& Power & Mean($N$) & SD($N$)
& Power & Mean($N$) & SD($N$) \\
\midrule

\multirow{6}{*}{$0$}
& OS
& 0.74581 & 67.49 & 31.35
& 0.77726 & 67.43 & 21.67
& 0.80438 & 67.31 & 9.30 \\

& UN
& 0.73800 & 63.19 & 31.19
& 0.77044 & 63.27 & 20.92
& 0.79948 & 63.47 & 8.61 \\

& Robust ($k=1.5$)
& 0.76936 & 74.68 & 38.56
& 0.78792 & 71.11 & 26.02
& 0.80667 & 68.15 & 10.89 \\

& Robust ($k=1.345$)
& 0.76879 & 76.46 & 41.11
& 0.78883 & 71.99 & 27.54
& 0.80809 & 68.35 & 11.40 \\

& Robust ($k=1.2$)
& 0.77150 & 78.52 & 44.35
& 0.79110 & 73.06 & 29.45
& 0.80809 & 68.61 & 12.02 \\

\midrule

\multirow{6}{*}{$0.1$}
& OS
& 0.66734 & 114.83 & 121.53
& 0.71902 & 114.59 & 87.32
& 0.79982 & 114.32 & 38.80 \\

& UN
& 0.65577 & 110.39 & 122.13
& 0.71148 & 110.44 & 87.42
& 0.79780 & 110.39 & 38.60 \\

& Robust ($k=1.5$)
& 0.64524 & 88.77 & 55.94
& 0.64310 & 81.07 & 32.39
& 0.64312 & 76.18 & 13.06 \\

& Robust ($k=1.345$)
& 0.64548 & 89.08 & 54.86
& 0.64436 & 81.12 & 33.13
& 0.64093 & 75.78 & 13.43 \\

& Robust ($k=1.2$)
& 0.64622 & 90.12 & 56.10
& 0.64308 & 81.56 & 34.53
& 0.63933 & 75.56 & 13.98 \\

\midrule

\multirow{6}{*}{$0.2$}
& OS
& 0.64165 & 160.94 & 163.19
& 0.70450 & 161.80 & 119.00
& 0.79153 & 161.22 & 53.01 \\

& UN
& 0.63251 & 156.32 & 164.31
& 0.70071 & 157.59 & 119.28
& 0.79044 & 157.25 & 52.83 \\

& Robust ($k=1.5$)
& 0.57516 & 107.49 & 79.42
& 0.56375 & 93.37 & 40.99
& 0.54707 & 85.85 & 15.67 \\

& Robust ($k=1.345$)
& 0.57425 & 105.45 & 73.86
& 0.55688 & 92.18 & 40.32
& 0.54233 & 84.61 & 15.76 \\

& Robust ($k=1.2$)
& 0.56926 & 104.76 & 71.91
& 0.55433 & 91.75 & 40.81
& 0.53767 & 83.73 & 16.14 \\

\midrule

\multirow{6}{*}{$0.3$}
& OS
& 0.64115 & 207.61 & 194.56
& 0.70502 & 208.39 & 142.92
& 0.78891 & 208.63 & 63.91 \\

& UN
& 0.63205 & 202.79 & 196.37
& 0.69801 & 204.21 & 143.78
& 0.78529 & 204.65 & 63.73 \\

& Robust ($k=1.5$)
& 0.53977 & 131.40 & 108.12
& 0.51520 & 108.99 & 52.73
& 0.49323 & 97.76 & 18.83 \\

& Robust ($k=1.345$)
& 0.53182 & 126.15 & 98.15
& 0.50559 & 105.85 & 49.74
& 0.48398 & 95.27 & 18.46 \\

& Robust ($k=1.2$)
& 0.52466 & 122.95 & 92.56
& 0.49983 & 104.01 & 48.67
& 0.47786 & 93.46 & 18.56 \\

\bottomrule
\end{tabular}
}
\end{table}

For continuous outcomes, all methods generally yielded empirical type~I error rates close to the nominal level across the scenarios considered, although small finite-sample deviations were observed in some settings. In the absence of contamination, the proposed method and the OS method controlled the type~I error rate satisfactorily, whereas the UN method showed type~I error inflation. The behavior of OS and UN methods is consistent with previous findings reported by \cite{FriedeKieser:2013} and \cite{MuetzeFriede:2021}.
When no contamination was present, the proposed method achieved higher power than OS and UN methods under $\sigma_1=\sigma_2=1$. However, under unequal variances ($\sigma_1=1, \sigma_2=2$), the proposed method showed some loss of power. When contamination was introduced, more pronounced differences emerged. The OS and UN methods exhibited substantial increases in both the mean and the variability of the final sample size, whereas the proposed method markedly reduced both quantities, indicating greater stability of the sample size re-estimation procedure. This advantage was most evident when the internal pilot sample size was small and became less pronounced as the pilot size increased. 
With respect to the relative efficiency ratio, the proposed method often showed more favorable sample-size-to-power trade-offs than OS method in both equal-variance and unequal-variance settings in the presence of contamination.
For continuous outcomes, the main advantage of the proposed method lies in mitigating sample size inflation and instability under contamination.

\subsection{Results for count outcome}

As described in the previous section, the sample size for the control group is determined so that the corresponding test attains a one-sided type I error rate of $\alpha$ under the null hypothesis $H_0: \log\lambda_1-\log\lambda_2=0$, and power $1-\beta$ to detect an alternative corresponding to $\delta_1=\log(0.75)$.

We assume that the true data-generating mechanism is given by
\begin{align*}
X_{i} \sim (1-\varepsilon)\mathrm{NB}(\lambda_1 t, \ \phi_1) + \varepsilon \mathrm{NB}(\lambda_1 t, \ \phi_o=2), \qquad Y_{j} \sim \mathrm{NB}(\lambda_2 t, \ \phi_2),
\end{align*}
where $\mathrm{NB}(\lambda_1 t, \ \phi_o=2)$ represents the contaminating distribution that generates outliers. A balanced design with equal follow-up time $t=1$ is considered.

\begin{table}[htbp]
\centering
\footnotesize
\caption{Simulation under the null hypothesis ($\lambda_1=\lambda_2=2$, $\phi_1=\phi_2=0.6$)}
\label{tab:combined_type1_sample_h0_equal}
\resizebox{\textwidth}{!}{
\begin{tabular}{llccccccccc}
\toprule
& & \multicolumn{3}{c}{$n_1=5$}
& \multicolumn{3}{c}{$n_1=10$}
& \multicolumn{3}{c}{$n_1=50$} \\
\cmidrule(lr){3-5} \cmidrule(lr){6-8} \cmidrule(lr){9-11}
$\epsilon$ & Method
& Type I & Mean($N$) & SD($N$)
& Type I & Mean($N$) & SD($N$)
& Type I & Mean($N$) & SD($N$) \\
\midrule

\multirow{6}{*}{$0$}
& OS
& 0.02511 & 216.88 & 125.62
& 0.02490 & 211.83 & 83.26
& 0.02539 & 211.09 & 35.10 \\

& UN
& 0.02665 & 209.79 & 127.93
& 0.02570 & 203.25 & 80.29
& 0.02591 & 207.53 & 34.79 \\

& Robust ($k=1.5$)
& 0.02521 & 218.67 & 111.99
& 0.02499 & 199.63 & 65.80
& 0.02528 & 188.35 & 25.14 \\

& Robust ($k=1.345$)
& 0.02590 & 227.87 & 113.74
& 0.02478 & 209.00 & 67.95
& 0.02580 & 197.37 & 26.20 \\

& Robust ($k=1.2$)
& 0.02469 & 241.69 & 118.04
& 0.02593 & 222.87 & 72.29
& 0.02522 & 211.04 & 27.94 \\

\midrule

\multirow{6}{*}{$0.1$}
& OS
& 0.02642 & 226.02 & 132.02
& 0.02684 & 221.39 & 88.27
& 0.02611 & 221.08 & 37.25 \\

& UN
& 0.02738 & 218.67 & 134.75
& 0.02697 & 212.35 & 85.18
& 0.02650 & 217.41 & 36.91 \\

& Robust ($k=1.5$)
& 0.02612 & 225.69 & 117.01
& 0.02753 & 205.78 & 68.92
& 0.02720 & 193.79 & 25.80 \\

& Robust ($k=1.345$)
& 0.02717 & 235.05 & 118.52
& 0.02725 & 215.33 & 70.83
& 0.02657 & 202.98 & 26.86 \\

& Robust ($k=1.2$)
& 0.02625 & 249.20 & 122.45
& 0.02626 & 229.64 & 74.80
& 0.02662 & 217.00 & 28.65 \\

\midrule

\multirow{6}{*}{$0.2$}
& OS
& 0.02745 & 235.64 & 137.82
& 0.02854 & 231.57 & 93.61
& 0.02755 & 231.23 & 39.26 \\

& UN
& 0.02802 & 227.90 & 140.73
& 0.02712 & 222.01 & 90.34
& 0.02724 & 227.44 & 38.90 \\

& Robust ($k=1.5$)
& 0.02856 & 233.01 & 121.31
& 0.02726 & 212.31 & 73.13
& 0.02855 & 199.25 & 26.38 \\

& Robust ($k=1.345$)
& 0.02810 & 242.51 & 122.55
& 0.02853 & 222.02 & 74.76
& 0.02705 & 208.61 & 27.45 \\

& Robust ($k=1.2$)
& 0.02901 & 256.99 & 126.02
& 0.02817 & 236.69 & 78.26
& 0.02863 & 222.95 & 29.27 \\

\midrule

\multirow{6}{*}{$0.3$}
& OS
& 0.02882 & 245.82 & 144.98
& 0.02944 & 242.50 & 98.27
& 0.02880 & 241.70 & 41.33 \\

& UN
& 0.02946 & 237.87 & 148.23
& 0.02902 & 232.39 & 94.67
& 0.02964 & 237.79 & 40.95 \\

& Robust ($k=1.5$)
& 0.02920 & 240.52 & 127.21
& 0.02910 & 219.22 & 76.13
& 0.02925 & 204.90 & 26.98 \\

& Robust ($k=1.345$)
& 0.02854 & 250.06 & 128.08
& 0.02841 & 229.16 & 77.58
& 0.02907 & 214.45 & 28.07 \\

& Robust ($k=1.2$)
& 0.02868 & 264.80 & 131.02
& 0.02871 & 244.26 & 80.74
& 0.02934 & 229.16 & 29.94 \\

\bottomrule
\end{tabular}
}
\end{table}

\begin{table}[htbp]
\centering
\footnotesize
\caption{Simulation under the alternative hypothesis ($\lambda_1=1.5$, $\lambda_2=2$, $\phi_1=\phi_2=0.6$)}
\label{tab:combined_results_equal}
\resizebox{\textwidth}{!}{
\begin{tabular}{llccccccccc}
\toprule
& & \multicolumn{3}{c}{$n_1=5$}
& \multicolumn{3}{c}{$n_1=10$}
& \multicolumn{3}{c}{$n_1=50$} \\
\cmidrule(lr){3-5} \cmidrule(lr){6-8} \cmidrule(lr){9-11}
$\epsilon$ & Method
& Power & Mean($N$) & SD($N$)
& Power & Mean($N$) & SD($N$)
& Power & Mean($N$) & SD($N$) \\
\midrule

\multirow{6}{*}{$0$}
& OS
& 0.74179 & 239.26 & 139.87
& 0.76712 & 232.12 & 93.34
& 0.79856 & 229.86 & 39.08 \\

& UN
& 0.73370 & 236.02 & 147.44
& 0.75528 & 222.38 & 90.12
& 0.79120 & 223.48 & 38.14 \\

& Robust ($k=1.5$)
& 0.76079 & 241.02 & 128.10
& 0.75834 & 218.71 & 76.86
& 0.75656 & 203.91 & 27.47 \\

& Robust ($k=1.345$)
& 0.77614 & 250.98 & 128.96
& 0.77584 & 228.89 & 78.31
& 0.77319 & 213.69 & 28.70 \\

& Robust ($k=1.2$)
& 0.79670 & 265.68 & 131.63
& 0.79892 & 244.14 & 81.51
& 0.80181 & 228.91 & 30.71 \\

\midrule

\multirow{6}{*}{$0.1$}
& OS
& 0.74082 & 248.07 & 145.56
& 0.76545 & 240.74 & 98.17
& 0.79792 & 238.88 & 40.86 \\

& UN
& 0.72955 & 245.08 & 154.67
& 0.74995 & 230.70 & 95.30
& 0.78942 & 232.30 & 39.99 \\

& Robust ($k=1.5$)
& 0.75628 & 248.08 & 132.43
& 0.75180 & 224.92 & 81.10
& 0.74798 & 209.21 & 28.09 \\

& Robust ($k=1.345$)
& 0.77146 & 258.18 & 133.04
& 0.77043 & 235.23 & 82.21
& 0.76896 & 219.18 & 29.36 \\

& Robust ($k=1.2$)
& 0.79256 & 273.23 & 135.35
& 0.79291 & 250.79 & 84.86
& 0.79310 & 234.73 & 31.46 \\

\midrule

\multirow{6}{*}{$0.2$}
& OS
& 0.73913 & 255.96 & 150.92
& 0.76608 & 250.24 & 102.26
& 0.79813 & 247.57 & 42.59 \\

& UN
& 0.72895 & 253.62 & 161.87
& 0.75192 & 239.73 & 99.39
& 0.79026 & 240.81 & 41.78 \\

& Robust ($k=1.5$)
& 0.75183 & 254.68 & 137.32
& 0.74823 & 231.61 & 84.88
& 0.74265 & 214.30 & 28.66 \\

& Robust ($k=1.345$)
& 0.76455 & 264.87 & 137.59
& 0.76488 & 242.09 & 85.75
& 0.76029 & 224.51 & 29.94 \\

& Robust ($k=1.2$)
& 0.78589 & 280.09 & 139.40
& 0.78725 & 257.94 & 87.97
& 0.78976 & 240.38 & 32.01 \\

\midrule

\multirow{6}{*}{$0.3$}
& OS
& 0.73594 & 265.72 & 156.78
& 0.76383 & 258.51 & 105.81
& 0.79469 & 256.69 & 44.41 \\

& UN
& 0.72653 & 263.48 & 167.81
& 0.75017 & 247.80 & 103.60
& 0.78795 & 249.75 & 43.68 \\

& Robust ($k=1.5$)
& 0.74876 & 262.36 & 142.27
& 0.74161 & 237.21 & 87.79
& 0.73569 & 219.66 & 29.37 \\

& Robust ($k=1.345$)
& 0.76202 & 272.57 & 142.32
& 0.75895 & 247.81 & 88.42
& 0.75438 & 230.06 & 30.63 \\

& Robust ($k=1.2$)
& 0.78407 & 288.06 & 143.70
& 0.78305 & 263.92 & 90.21
& 0.78425 & 246.25 & 32.72 \\

\bottomrule
\end{tabular}
}
\end{table}

For count outcomes, the simulation results showed that the empirical type~I error rates were close to the nominal level across most scenarios, although mild inflation was observed in some contaminated settings. In the absence of contamination, the proposed method and the OS method provided satisfactory type~I error control, whereas the UN method exhibited type~I error inflation. The behavior of OS and UN methods is consistent with previous findings reported by \cite{Schneideretal:2013b}.
Compared with the OS and UN methods, the proposed method often reduced both the mean and the variability of the final sample size, particularly when the internal pilot sample size was moderate or large. Under stronger contamination or unequal dispersion between groups, however, the proposed method generally involved a trade-off between improved stability of the final sample size and some loss of power. When the dispersion was equal between groups, the proposed method with a smaller tuning constant achieved higher power than OS and UN methods. In terms of the relative efficiency ratio, the proposed method was generally more favorable than OS method. For count outcomes, robust blinded sample size re-estimation can improve design stability under model deviations.

\section{Clinical trial application}\label{sec-4}

\begin{figure}
  \centering
  \includegraphics[width=400pt]{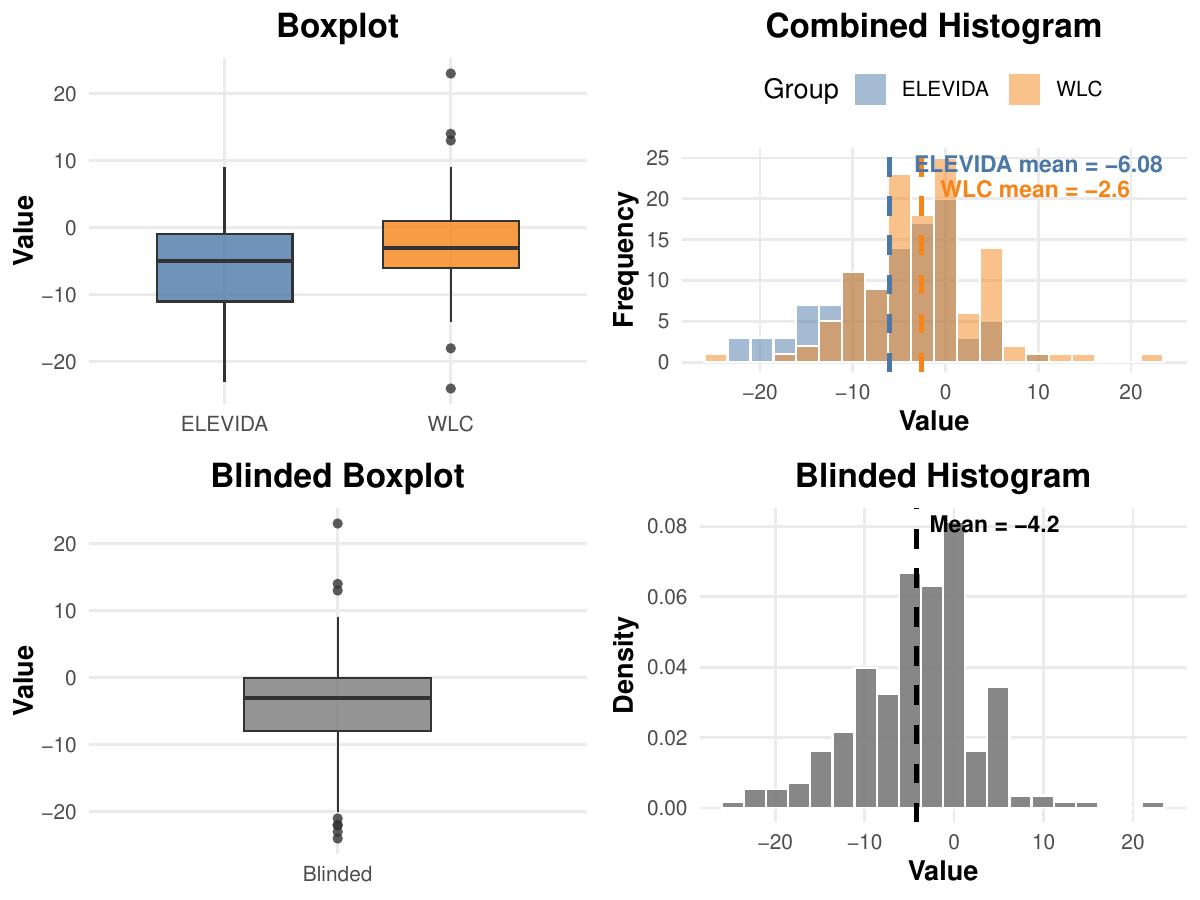}\\
  \caption{Empirical distribution of the completed dataset (the blinded boxplot and blinded histogram obtained by ignoring the group labels)}\label{f1}
\end{figure}

In this section, we reanalyze data from a completed clinical trial using a sample size re-estimation procedure to illustrate the practical utility of the proposed method.
\cite{Poettgenetal:2018} evaluated the effectiveness of ELEVIDA, a self-guided, interactive, online fatigue management programme, in reducing fatigue among patients with multiple sclerosis, compared with a wait-list control group (WLC).
The primary endpoint was the Chalder Fatigue Scale, assessed at baseline and again at Week~12.
The scale consists of 11 items, each scored from 0 to 3, yielding a total score ranging from 0 to 33, with higher scores indicating greater fatigue severity. 
To provide an overview of the dataset, Figure~\ref{f1} displays the empirical distribution of the complete data, including a blinded boxplot and a blinded histogram with group labels ignored. The blinded boxplot suggests the presence of extreme observations in the pooled data.
After unblinding, the group-specific boxplot indicates that several of these observations appear in the WLC group.

All data used in this reanalysis were randomly sampled from the fully observed dataset reported by \cite{Poettgenetal:2018}.
For the internal pilot study, equal numbers of participants, denoted by $n_1$, were allocated to the treatment group (ELEVIDA) and the control group (WLC).
A one-sided hypothesis test was considered, with the nominal significance level set at $\alpha=0.025$ and the target type~II error rate set at $\beta=0.2$.
The minimal clinically important difference was taken to be 1.4. 

The outcome of interest is the change in Chalder Fatigue Scale score from baseline to Week~12, defined as the Week~12 score minus the baseline score, with negative values indicating an improvement in fatigue symptoms. The estimand is the between-group difference in mean changes, defined as $\mu_1-\mu_2$, where $\mu_1$ and $\mu_2$ denote the population mean changes in the wait-list control (WLC) and ELEVIDA treatment groups, respectively. Specifically, we test the null hypothesis $H_0:\mu_1=\mu_2$ against the one-sided alternative $H_1:\mu_1>\mu_2$.
Initial samples of $n_1=5$, $10$, and $50$ patients were randomly selected from each group, after which sample size re-estimation was performed.
We compared several variance estimators for the re-estimation.
With the exception of the unblinded estimator (UN), all methods were blinded and did not use treatment assignment information.

Table~\ref{tab:combined_variance_sample_size} illustrates the impact of different variance estimators on the final sample size. For small internal pilot samples ($n_1=5$ or $10$), the OS and UN methods yield substantially larger variance estimates and, consequently, larger final sample sizes than the proposed robust method. These differences become less pronounced as the pilot sample size increases to $n_1=50$. In this application, the proposed robust estimator appears less sensitive to extreme observations, resulting in more moderate sample size estimates, particularly when the internal pilot sample size is small.

\begin{table}[htbp]
\centering
\caption{Estimated variance and final sample size per group $N$ under different methods}
\label{tab:combined_variance_sample_size}
\resizebox{\textwidth}{!}{
\begin{tabular}{llcccccc}
\toprule
& & OS & UN & Robust ($k=1.5$) & Robust ($k=1.345$) & Robust ($k=1.2$)  \\
\midrule
\multirow{2}{*}{$n_1=5$}
& Estimated variance        & 58.18 & 59.40 & 29.33 & 26.50 & 25.91  \\
& $N$  & 466   & 476   & 235   & 213   & 208   \\
\midrule
\multirow{2}{*}{$n_1=10$}
& Estimated variance        & 63.46 & 49.21 & 35.42 & 35.41 & 36.78  \\
& $N$  & 509   & 395   & 284   & 284   & 295    \\
\midrule
\multirow{2}{*}{$n_1=50$}
& Estimated variance        & 48.88 & 48.17 & 45.58 & 46.36 & 47.11  \\
& $N$  & 392   & 386   & 366   & 372   & 378   \\
\bottomrule
\end{tabular}
}
\end{table}

\section{Conclusions}\label{sec-5}

In this paper, we develop a unified framework for robust blinded sample size re-estimation in clinical trials, encompassing several existing approaches. We derive a general sample size formula based on asymptotically linear estimators and introduce a reparameterization principle for defining nuisance parameters in terms of a prespecified effect measure and a blinded summary statistic. From a theoretical perspective, we establish that, under suitable regularity conditions, the final sample size and its expectation are asymptotically equivalent to the target sample size, while the nominal type~I error level is asymptotically preserved.

For nuisance parameter estimation, we advocate robust $Z$-estimators as an alternative to conventional likelihood-based methods. Likelihood-based estimators may be sensitive to extreme observations and model misspecification, potentially leading to unstable nuisance parameter estimates and increased variability in the final sample size, particularly when the internal pilot sample is small. In contrast, robust $Z$-estimators can reduce the influence of extreme observations and improve the stability of sample size re-estimation under contamination. These properties make robust estimation a promising approach for adaptive clinical trial designs.

In the absence of contamination, the robust procedure based on Huber's Proposal 2 can achieve power closer to the nominal target in certain settings with equal variances for continuous outcomes or equal dispersion parameters for count outcomes. Under contamination or model misspecification, the proposed robust methods can reduce the influence of extreme observations on nuisance parameter estimation, thereby improving the stability of the final sample size.

The performance of Huber's Proposal 2 depends on the choice of the tuning parameter $k$. In Section~\ref{sec-3}, we investigate three values of $k$ to assess its impact on the operating characteristics of the proposed procedure. Smaller values of $k$ impose stronger downweighting of extreme residuals, potentially improving robustness at the expense of efficiency under the working model. However, their effects on the final sample size and statistical power depend on the underlying outcome distribution and the degree of contamination. In confirmatory clinical trials, the tuning parameter should preferably be prespecified in the study protocol. Data-driven approaches to selecting $k$, such as cross-validation, may be explored in future research, provided that their effects on the operating characteristics of the design are carefully evaluated.


Several directions merit further investigation. One promising extension is to adapt the proposed framework to continuous monitoring (cf. \cite{FriedeMiller:2012}, \cite{XuFriede:2026}), in which nuisance parameters are repeatedly updated as additional data become available. Another direction is motivated by \cite{Maedaetal:2026}, who proposed using an upper confidence bound for the variance to mitigate the risk of underpowered studies. Building on this idea, future work could investigate the construction of confidence regions for the nuisance parameter vector $\bm{\eta}$ and determine the final sample size by maximizing the sample size planning function over these regions. Such an approach may provide a more conservative sample size re-estimation procedure, potentially reducing the risk of insufficient statistical power.

\section*{Acknowledgments}
We would like to thank the Editor, Associate Editor, and anonymous reviewers for their valuable comments and suggestions, which have improved the presentation of the manuscript.
The authors are grateful for funding by the DFG grant numbers FR 3070/4-1 and FR 3070/4-2.

\newpage

\appendix

\section*{Supplementary Materials}
\addcontentsline{toc}{section}{Supplementary Materials}

\setcounter{table}{0}
\setcounter{figure}{0}
\setcounter{equation}{0}

\renewcommand{\thetable}{S\arabic{table}}
\renewcommand{\thefigure}{S\arabic{figure}}
\renewcommand{\theequation}{S\arabic{equation}}

The Supplementary Material contains proofs of the theoretical results, additional simulation results, and several examples illustrating the applicability of the proposed framework across a variety of settings. 
We show that several existing procedures in the literature can be recovered as special cases and derive new robust sample size re-estimation procedures within the same unified framework. 
The R code used to reproduce the simulation results is available at https://github.com/xu19971997/A-Unified-Framework-for-Robust-Blinded-Sample-Size-Re-estimation-Based-on-Nuisance-Parameters.

\section{Proofs}

We impose the following regularity conditions on $n(\bm{\eta})$:
\begin{itemize}
\item[N1]
We assume that
\begin{align*}
n(\bm{\eta}_0)\rightarrow\infty,\quad n_1=n_1(n(\bm{\eta}_0))\rightarrow\infty,\quad \limsup(n_1/n(\bm{\eta}_0))<1,\quad n(\bm{\eta}_0)=O(n_1^l)
\end{align*}
where $l\geq 1$ is a fixed constant.

\item[N2] There exists a neighborhood $\mathcal U(\bm{\eta}_0)$ of $\bm{\eta}_0$ such that $n(\bm{\eta})$ is continuously differentiable on $\mathcal U(\bm{\eta}_0)$. Moreover, $n(\bm{\eta}_0)>0$, and there exists
a constant $C_1<\infty$ such that
\begin{align*}
\sup_{\bm{\eta}\in\mathcal U(\bm{\eta}_0)}
\frac{
\|\nabla n(\bm{\eta})\|
}{
n(\bm{\eta}_0)
}
\le C_1
\end{align*}
for all sufficiently large $n(\bm{\eta}_0)$.

\item[N3] There exist constants $p>1$ and $C_2<\infty$ such that
\begin{align*}
\sup_{n_1}\,
\mathbb{E}\left|
\frac{n(\widehat{\bm{\eta}})}{n(\bm{\eta}_0)}
\right|^p
\le C_2.
\end{align*}

\item[N4] The random sample size $N$ is assumed to be $\mathcal{G}_{m_1,n_1}$-measurable, where
\begin{align*}
\mathcal{G}_{m_1,n_1}:=\sigma\!(\{X_i\}_{i\le m_1},\{Y_j\}_{j\le n_1})
\end{align*}
denote the pilot sigma-field.

\item[N5]
There exists a fixed constant $\delta\in(0,1)$ such that
\begin{align*}
\sup_{m:\,|m/(rn(\bm\eta_0))-1|\le \delta}|\sqrt{m}\,(\widehat\theta_{1,m}-\theta_1)-\frac{1}{\sqrt m}\sum_{i=1}^m \varphi_1(X_i;\theta_1,\omega)| \overset{p}{\rightarrow}0,
\end{align*}
and
\begin{align*}
\sup_{n:\,|n/n(\bm\eta_0)-1|\le \delta}|\sqrt{n}\,(\widehat\theta_{2,n}-\theta_2)-\frac{1}{\sqrt n}\sum_{j=1}^n \varphi_2(Y_j;\theta_2,\omega)| \overset{p}{\rightarrow}0.
\end{align*}
\end{itemize}

Condition N1 is similar to those used in \cite{Liu:1997} and \cite{Hall:1981}, and can be regarded as a standard assumption in this context.
Condition N2 is a local relative smoothness condition for the sample size function.
Condition N3 guarantees the uniform integrability. 
Condition N4 formalizes that the sample size re-estimation rule depends only on the internal pilot data and does not use future observations.
Condition N5 guarantees that the fixed-index asymptotic linear representations continue to hold after the deterministic indices are replaced by the random indices $M$ and $N$.

The following regularity conditions are assumed for the $Z$-estimator.
\begin{itemize}
\item[R1] The parameter space $\Theta$ is compact and contains $\bm{\eta}_0$.

\item[R2] The population estimating function
\begin{align*}
\Psi_0(\bm{\eta})
=
\frac{r}{r+1}
\mathbb{E}_{\theta_1,\omega}(\Psi(\bm{\eta};X))
+
\frac{1}{r+1}
\mathbb{E}_{\theta_2,\omega}(\Psi(\bm{\eta};Y))
\end{align*}
satisfies
\begin{align*}
\Psi_0(\bm{\eta}_0)=0,
\end{align*}
and $\bm{\eta}_0$ is the unique solution of $\Psi_0(\bm{\eta})=0$ over $\Theta$.

\item[R3] For each $z$, $\Psi(\bm{\eta};z)$ is continuous in $\bm{\eta}$. Moreover,
\begin{align*}
\mathbb{E}_{\theta_1,\omega}\sup_{\bm{\eta}\in\Theta}\|\Psi(\bm{\eta};X)\|<\infty,
\qquad
\mathbb{E}_{\theta_2,\omega}\sup_{\bm{\eta}\in\Theta}\|\Psi(\bm{\eta};Y)\|<\infty.
\end{align*}

\item[R4] A uniform law of large numbers holds for the two-sample empirical
estimating function:
\begin{align*}
\sup_{\bm{\eta}\in\Theta}
\left\|
\frac{1}{m_1}\sum_{i=1}^{m_1}\Psi(\bm{\eta};X_i)
-
\mathbb{E}_{\theta_1,\omega}(\Psi(\bm{\eta};X))
\right\|
\xrightarrow{p}0,
\end{align*}
and
\begin{align*}
\sup_{\bm{\eta}\in\Theta}
\left\|
\frac{1}{n_1}\sum_{j=1}^{n_1}\Psi(\bm{\eta};Y_j)
-
\mathbb{E}_{\theta_2,\omega}(\Psi(\bm{\eta};Y))
\right\|
\xrightarrow{p}0.
\end{align*}
\end{itemize}

Assumptions R1--R4 are standard regularity conditions in the general theory of $Z$-estimators; see, for example, \cite{VanderVaart:2000}.

\subsection{Proof of Lemma 1}

By Assumption R4 and the fixed allocation ratio $m_1/n_1=r$, we have
\begin{align*}
\sup_{\bm{\eta}\in\Theta}
\left\|
\frac{1}{n_p}\sum_{k=1}^{n_p}\Psi(\bm{\eta};Z_k)
-
\Psi_0(\bm{\eta})
\right\|
\xrightarrow{p}0,
\end{align*}
where
\begin{align*}
\Psi_0(\bm{\eta})
=
\frac{r}{r+1}
\mathbb{E}_{\theta_1,\omega}(\Psi(\bm{\eta};X))
+
\frac{1}{r+1}
\mathbb{E}_{\theta_2,\omega}(\Psi(\bm{\eta};Y)).
\end{align*}
Together with compactness of $\Theta$, continuity of $\Psi_0$, and uniqueness of
the zero $\bm{\eta}_0$ in Assumption R2, the standard consistency theorem for $Z$-estimators yields $\widehat{\bm{\eta}}\xrightarrow{p}\bm{\eta}_0$.

Let $\mathcal U(\bm{\eta}_0)$ be the neighborhood specified in Assumption N2.
Since $\widehat{\bm{\eta}}\xrightarrow{p}\bm{\eta}_0$,
we have $\mathbb{P}(\widehat{\bm{\eta}}\in\mathcal U(\bm{\eta}_0))\to1$.
On the event
$\{\widehat{\bm{\eta}}\in\mathcal U(\bm{\eta}_0)\}$, the mean value theorem gives
\begin{align*}
\left|
\frac{n(\widehat{\bm{\eta}})}
{n(\bm{\eta}_0)}
-1
\right|
=
\frac{
|n(\widehat{\bm{\eta}})-n(\bm{\eta}_0)|
}{
n(\bm{\eta}_0)
}
\le
\sup_{\bm{\eta}\in\mathcal U(\bm{\eta}_0)}
\frac{
\|\nabla n(\bm{\eta})\|
}{
n(\bm{\eta}_0)
}
\,
\|\widehat{\bm{\eta}}-\bm{\eta}_0\|.
\end{align*}
By Assumption N2, the supremum on the right-hand side is bounded by a constant $C_1<\infty$ for all sufficiently large $n(\bm{\eta}_0)$. Hence
\begin{align*}
\left|
\frac{n(\widehat{\bm{\eta}})}
{n(\bm{\eta}_0)}
-1
\right|
\le
C_1\|\widehat{\bm{\eta}}-\bm{\eta}_0\|
=
o_p(1).
\end{align*}
Therefore,
\begin{align*}
\frac{n(\widehat{\bm{\eta}})}
{n(\bm{\eta}_0)}
\xrightarrow{p}1.
\end{align*}

Since
\begin{align*}
n(\widehat{\bm{\eta}})
<
\lfloor n(\widehat{\bm{\eta}})\rfloor+1
\le
n(\widehat{\bm{\eta}})+1,
\end{align*}
and $n(\bm{\eta}_0)\to\infty$, it follows that
\begin{align*}
\frac{\lfloor n(\widehat{\bm{\eta}})\rfloor+1}{n(\bm{\eta}_0)}
\overset{p}{\to}1.
\end{align*}
Moreover, by Assumption N1, there exists $\delta\in(0,1)$ such that
\begin{align*}
\frac{n_1}{n(\bm{\eta}_0)}\le 1-\delta
\end{align*}
for all sufficiently large $n(\bm{\eta}_0)$. Therefore,
\begin{align*}
\frac{N}{n(\bm{\eta}_0)}
=
\max\left\{
\frac{\lfloor n(\widehat{\bm{\eta}})\rfloor+1}{n(\bm{\eta}_0)},
\frac{n_1}{n(\bm{\eta}_0)}
\right\}
\overset{p}{\to}1.
\end{align*}

It remains to justify convergence of expectations. 
Using $\lfloor x\rfloor+1\le x+1$, we obtain
\begin{align*}
\frac{N}{n(\bm{\eta}_0)}
\le
\max\left\{
\frac{n(\widehat{\bm\eta})}{n(\bm{\eta}_0)}
+
\frac{1}{n(\bm{\eta}_0)},
\frac{n_1}{n(\bm{\eta}_0)}
\right\}.
\end{align*}
Since both terms are nonnegative, $\max(a,b)\le a+b$ gives
\begin{align*}
\frac{N}{n(\bm{\eta}_0)}
\le
\frac{n(\widehat{\bm\eta})}{n(\bm{\eta}_0)}
+
\frac{1}{n(\bm{\eta}_0)}
+
\frac{n_1}{n(\bm{\eta}_0)}.
\end{align*}
By Assumption N1, $n_1/n(\bm{\eta}_0)$ is bounded for all sufficiently large $n(\bm{\eta}_0)$, and $1/n(\bm{\eta}_0)\to0$. Hence there exists a constant $K<\infty$ such that, for all sufficiently large \(n(\eta_0)\),
\begin{align*}
\frac{1}{n(\bm{\eta}_0)}+\frac{n_1}{n(\bm{\eta}_0)}\le K.
\end{align*}
Therefore,
\begin{align*}
\frac{N}{n(\bm{\eta}_0)}
\le
\frac{n(\widehat{\bm\eta})}{n(\bm{\eta}_0)}+K.
\end{align*}
For the $p>1$ in Assumption N3, it follows from Jensen's inequality $(a+b)^p\le 2^{p-1}(a^p+b^p)$ that
\begin{align*}
\mathbb{E}\left[
\left|
\frac{N}{n(\bm{\eta}_0)}
\right|^p
\right]
\le
2^{p-1}
\mathbb{E}\left[
\left|
\frac{n(\widehat{\bm\eta})}{n(\bm{\eta}_0)}
\right|^p
\right]
+
2^{p-1}K^p.
\end{align*}
By Assumption N3,
\begin{align*}
\sup_{n_1}
\mathbb{E}\left[
\left|
\frac{N}{n(\bm{\eta}_0)}
\right|^p
\right]
<\infty .
\end{align*}
Thus $\{N/n(\bm{\eta}_0)\}$ is uniformly integrable. Since we have already shown that
\begin{align*}
\frac{N}{n(\bm{\eta}_0)}\xrightarrow{p}1,
\end{align*}
uniform integrability implies
\begin{align*}
\mathbb{E}\left(\frac{N}{n(\bm{\eta}_0)}\right)\to1.
\end{align*}

This completes the proof.

\begin{flushright}
$\square$
\end{flushright}

\subsection{Proof of Lemma 2}

Let $A=n(\widehat{\bm{\eta}})$ and define the final sample size by
\[
    N=\max\{\lfloor A\rfloor+1,n_1\}.
\]

Set
\[
    Y=\max\{A,n_1\},
    \qquad
    D=N-Y.
\]
Since $Y\le A+n_1$ and $N\le A+n_1+1$, both $Y$ and $N$
have finite second moments. Moreover, because $n_1$ is an
integer,
\[
    D=
    \begin{cases}
        0, & A<n_1,\\[1mm]
        \lfloor A\rfloor+1-A, & A\ge n_1.
    \end{cases}
\]
Consequently,
\begin{equation}
\label{eq:rounding-error-range}
    0\le D\le 1
    \qquad\text{almost surely}.
\end{equation}

We first show that imposing the lower bound $n_1$ does
not increase the variance of the continuous planning
quantity. Let $A'$ be an independent copy of $A$, and
let $Y'=\max\{A',n_1\}$. The function
$h(x)=\max\{x,n_1\}$ is $1$-Lipschitz, so
\[
    |Y-Y'|
    =
    |h(A)-h(A')|
    \le |A-A'|.
\]
Using the independent-copy identity for variance gives
\begin{equation}
\label{eq:lower-truncation-variance}
\begin{aligned}
    \operatorname{Var}(Y)
    &=
    \frac{1}{2}\mathbb{E}\{(Y-Y')^2\}
    \le
    \frac{1}{2}\mathbb{E}\{(A-A')^2\}
    =
    \operatorname{Var}(A).
\end{aligned}
\end{equation}

Next, \eqref{eq:rounding-error-range} implies $D^2\le D$
almost surely. Hence,
\begin{equation}
\label{eq:rounding-error-variance}
\begin{aligned}
    \operatorname{Var}(D)
    &=
    \mathbb{E}(D^2)-\{\mathbb{E}(D)\}^2
    \le
    \mathbb{E}(D)\{1-\mathbb{E}(D)\}
    \le
    \frac{1}{4}.
\end{aligned}
\end{equation}

Since $N=Y+D$, the Cauchy--Schwarz inequality yields
\[
\begin{aligned}
    \operatorname{Var}(N)
    &=
    \operatorname{Var}(Y)+\operatorname{Var}(D)
    +2\operatorname{Cov}(Y,D)\\
    &\le
    \operatorname{Var}(Y)+\operatorname{Var}(D)
    +2\sqrt{\operatorname{Var}(Y)\operatorname{Var}(D)}\\
    &=
    \left\{
        \sqrt{\operatorname{Var}(Y)}
        +\sqrt{\operatorname{Var}(D)}
    \right\}^{2}.
\end{aligned}
\]
Taking square roots and applying
\eqref{eq:lower-truncation-variance} and
\eqref{eq:rounding-error-variance}, we obtain
\[
    \sqrt{\operatorname{Var}(N)}
    \le
    \sqrt{\operatorname{Var}(Y)}
    +\sqrt{\operatorname{Var}(D)}
    \le
    \sqrt{\operatorname{Var}(A)}+\frac{1}{2}.
\]

To establish optimality, fix any positive integer $n_1$
and let $0<\varepsilon<1$. Define $A_\varepsilon$ by
\[
    \mathbb{P}(A_\varepsilon=n_1+1-\varepsilon)
    =
    \mathbb{P}(A_\varepsilon=n_1+1)
    =
    \frac{1}{2}.
\]
Then
\[
    N_\varepsilon
    =
    \max\{\lfloor A_\varepsilon\rfloor+1,n_1\}
\]
satisfies
\[
    \mathbb{P}(N_\varepsilon=n_1+1)
    =
    \mathbb{P}(N_\varepsilon=n_1+2)
    =
    \frac{1}{2}.
\]
Therefore,
\[
    \operatorname{Var}(A_\varepsilon)
    =\frac{\varepsilon^2}{4},
    \qquad
    \operatorname{Var}(N_\varepsilon)
    =\frac{1}{4},
\]
and thus
\[
    \sqrt{\operatorname{Var}(N_\varepsilon)}
    -
    \sqrt{\operatorname{Var}(A_\varepsilon)}
    =
    \frac{1-\varepsilon}{2}
    \longrightarrow
    \frac{1}{2}
    \qquad\text{as }\varepsilon\downarrow0.
\]
Consequently, every constant smaller than $\frac{1}{2}$ fails
for a sufficiently small positive $\varepsilon$.
This proves optimality for each fixed $n_1$.

\begin{flushright}
$\square$
\end{flushright}

\subsection{Proof of Theorem 1}

By Lemma \ref{lemma-1}, under Assumptions N1--N5 and R1--R4, we have
\begin{align*}
\frac{N}{n(\bm\eta_0)}\overset{p}{\rightarrow}1.
\end{align*}
Since $n(\bm\eta_0)\to\infty$, it follows that $N\overset{p}{\rightarrow}\infty$.
Because $M= rN$, we also have
\begin{align*}
\frac{M}{r n(\bm\eta_0)}\overset{p}{\rightarrow}1.
\end{align*}

For \(m,n\geq 1\), define the remainder terms
\begin{align*}
R_{1,m}
=
\sqrt{m}(\widehat\theta_{1,m}-\theta_1)
-
\frac{1}{\sqrt{m}}
\sum_{i=1}^{m}\phi_1(X_i;\theta_1,\omega),
\end{align*}
and
\begin{align*}
R_{2,n}
=
\sqrt{n}(\widehat\theta_{2,n}-\theta_2)
-
\frac{1}{\sqrt{n}}
\sum_{j=1}^{n}\phi_2(Y_j;\theta_2,\omega).
\end{align*}

We next verify that the fixed-index asymptotic linear representations remain valid after replacing the deterministic indices by the random indices $M$ and $N$. We only prove the statement for $R_{2,N}$, since the argument for $R_{1,M}$ is entirely analogous.

Let
\begin{align*}
S:=\left\{n:\left|\frac{n}{n(\bm\eta_0)}-1\right|\le \delta\right\}.
\end{align*}
Then, for any $\varepsilon>0$,
\begin{align*}
\{|R_{2,N}|>\varepsilon\}
=
\bigl(\{|R_{2,N}|>\varepsilon\}\cap \{N\in S\}\bigr)
\;\cup\;
\bigl(\{|R_{2,N}|>\varepsilon\}\cap \{N\notin S\}\bigr).
\end{align*}
Hence,
\begin{align*}
\mathbb P(|R_{2,N}|>\varepsilon)
\le
\mathbb P\bigl(\{|R_{2,N}|>\varepsilon\}\cap \{N\in S\}\bigr)
+
\mathbb P(N\notin S).
\end{align*}
Now, on the event $\{N\in S\}$, the random index $N$ belongs to the set over which the supremum $\sup_{n\in S}|R_{2,n}|$ is taken. Therefore,
\begin{align*}
\{|R_{2,N}|>\varepsilon\}\cap \{N\in S\}
\subseteq
\left\{\sup_{n\in S}|R_{2,n}|>\varepsilon\right\}.
\end{align*}
It follows that
\begin{align*}
\mathbb P(|R_{2,N}|>\varepsilon)
\le
\mathbb P\!\left(\sup_{n\in S}|R_{2,n}|>\varepsilon\right)
+
\mathbb P(N\notin S).
\end{align*}
By the definition of $S$,
\begin{align*}
\{N\notin S\}
=
\left\{\left|\frac{N}{n(\bm\eta_0)}-1\right|>\delta\right\},
\end{align*}
and thus
\begin{align*}
\mathbb P(|R_{2,N}|>\varepsilon)
\le
\mathbb P\!\left(
\sup_{n:\,|n/n(\bm\eta_0)-1|\le\delta}|R_{2,n}|>\varepsilon
\right)
+
\mathbb P\!\left(
\left|\frac{N}{n(\bm\eta_0)}-1\right|>\delta
\right).
\end{align*}

By Assumption N5,
\begin{align*}
\sup_{n:\,|n/n(\bm\eta_0)-1|\le\delta}|R_{2,n}|
\overset{p}{\rightarrow}0,
\end{align*}
so the first probability on the right-hand side converges to zero. Moreover, since
\begin{align*}
\frac{N}{n(\bm\eta_0)}\overset{p}{\longrightarrow}1,
\end{align*}
the second probability also converges to zero. Therefore, for every $\varepsilon>0$,
\begin{align*}
\mathbb P(|R_{2,N}|>\varepsilon)\rightarrow 0,
\end{align*}
which proves that
\begin{align*}
R_{2,N}\overset{p}{\rightarrow}0.
\end{align*}

Similarly,
\begin{align*}
R_{1,M}\overset{p}{\rightarrow}0.
\end{align*}

Therefore,
\begin{align*}
\sqrt{M}\,(\widehat\theta_{1,M}-\theta_1)
=
\frac{1}{\sqrt M}\sum_{i=1}^M \varphi_1(X_i;\theta_1,\omega)+o_p(1),
\end{align*}
and
\begin{align*}
\sqrt{N}\,(\widehat\theta_{2,N}-\theta_2)
=
\frac{1}{\sqrt N}\sum_{j=1}^N \varphi_2(Y_j;\theta_2,\omega)+o_p(1).
\end{align*}

Since $N$ is measurable with respect to the pilot sigma-field $\mathcal G_{m_1,n_1}$, and the future observations are independent of the pilot data by Assumption N4, the random indices $M$ and $N$ depend only on the pilot sample.
Since $M/(rn(\bm\eta_0))\overset{p}{\to}1$ and $N/n(\bm\eta_0)\overset{p}{\to}1$, Anscombe's theorem (e.g. \cite[Theorem 3.1]{Gut:2009}) yields
\begin{align*}
\frac{1}{\sqrt M}\sum_{i=1}^M \varphi_1(X_i;\theta_1,\omega)
\overset{d}{\rightarrow} \mathcal{N}(0,\tau_1^2),
\qquad
\frac{1}{\sqrt N}\sum_{j=1}^N \varphi_2(Y_j;\theta_2,\omega)
\overset{d}{\rightarrow} \mathcal{N}(0,\tau_2^2).
\end{align*}

Fix arbitrary $(a,b)\in\mathbb R^2$. Since $M=rN$, define
\begin{align*}
\xi_k
:=
\frac{a}{\sqrt r} \left(\sum_{\ell=1}^r
\varphi_1\!\left(X_{r(k-1)+\ell};\theta_1,\omega\right) \right)
+
b\,\varphi_2(Y_k;\theta_2,\omega),
\qquad k\ge 1.
\end{align*}
Because the $X$-sample and $Y$-sample are independent and each sequence is i.i.d.,
$\{\xi_k\}_{k\ge1}$ is an i.i.d. sequence with
\begin{align*}
\mathbb{E}(\xi_k)=0,
\qquad
\mbox{Var}(\xi_k)=a^2\tau_1^2+b^2\tau_2^2.
\end{align*}
Moreover,
\begin{align*}
a\,\frac1{\sqrt M}\sum_{i=1}^M \varphi_1(X_i;\theta_1,\omega)
+
b\,\frac1{\sqrt N}\sum_{j=1}^N \varphi_2(Y_j;\theta_2,\omega)
=
\frac1{\sqrt N}\sum_{k=1}^N \xi_k.
\end{align*}
Since $N/n(\bm\eta_0)\overset{p}{\to}1$ and $n(\bm\eta_0)\to\infty$, Anscombe's theorem (e.g. \cite[Theorem 3.1]{Gut:2009}) for randomly
indexed partial sums gives
\begin{align*}
\frac1{\sqrt N}\sum_{k=1}^N \xi_k
\overset{d}{\rightarrow}
\mathcal{N}(0,a^2\tau_1^2+b^2\tau_2^2).
\end{align*}

The Cram\'{e}r--Wold device then implies
\begin{align*}
\left(
\frac{1}{\sqrt M}\sum_{i=1}^M \varphi_1(X_i;\theta_1,\omega),
\frac{1}{\sqrt N}\sum_{j=1}^N \varphi_2(Y_j;\theta_2,\omega)
\right)^{T}
\overset{d}{\rightarrow}
\mathcal{N}_2\!\left(
\begin{pmatrix}0\\0\end{pmatrix},
\begin{pmatrix}
\tau_1^2 & 0\\
0 & \tau_2^2
\end{pmatrix}
\right).
\end{align*}

Combining this joint convergence with the two random-index linear expansions above and Slutsky's theorem yields
\begin{align*}
\begin{pmatrix}
\sqrt{M}\,(\widehat\theta_{1,M}-\theta_1)\\[1mm]
\sqrt{N}\,(\widehat\theta_{2,N}-\theta_2)
\end{pmatrix}
\overset{d}{\rightarrow}
\mathcal{N}_2\!\left(
\begin{pmatrix}0\\0\end{pmatrix},
\begin{pmatrix}
\tau_1^2 & 0\\
0 & \tau_2^2
\end{pmatrix}
\right).
\end{align*}

Under the null hypothesis $H_0:\theta_1=\theta_2$, we may write
\begin{align*}
\sqrt{M+N}\,(\widehat\theta_{1,M}-\widehat\theta_{2,N})
&=
\sqrt{\frac{M+N}{M}}\,
\sqrt{M}(\widehat\theta_{1,M}-\theta_1)
-
\sqrt{\frac{M+N}{N}}\,
\sqrt{N}(\widehat\theta_{2,N}-\theta_2)\\
&=
\sqrt{1+\frac1r}\,
\sqrt{M}(\widehat\theta_{1,M}-\theta_1)
-
\sqrt{1+r}\,
\sqrt{N}(\widehat\theta_{2,N}-\theta_2).
\end{align*}
Applying the continuous mapping theorem and Slutsky's theorem to the joint convergence above, we obtain
\begin{align*}
\sqrt{M+N}\,(\widehat\theta_{1,M}-\widehat\theta_{2,N})
\overset{d}{\rightarrow}
\mathcal{N}\!\left(
0,
\left(1+\frac1r\right)\tau_1^2+(1+r)\tau_2^2
\right).
\end{align*}

By the consistency of the final variance estimators,
\begin{align*}
\widehat\tau_{1,0}^2\overset{p}{\rightarrow}\tau_1^2,
\qquad
\widehat\tau_{2,0}^2\overset{p}{\rightarrow}\tau_2^2.
\end{align*}
Therefore, by Slutsky's theorem,
\begin{align*}
T_{M,N}
\overset{d}{\rightarrow}
\mathcal{N}(0,1),
\end{align*}
under the null hypothesis $H_0:\theta_1=\theta_2$.
This completes the proof.

\begin{flushright}
$\square$
\end{flushright}

\section{Examples}\label{ap-b}

In this section, we present several examples to illustrate the applicability of the proposed framework in a variety of settings. We show that several existing procedures in the literature can be recovered as special cases, and we also derive new robust sample size re-estimation procedures within this unified framework.

\begin{exam}\label{example-1}
(Normal distribution)

Consider $X_i\sim \mathcal{N}(\mu_1,\sigma^2)$ and $Y_j\sim \mathcal{N}(\mu_2,\sigma^2)$.
We are interested in testing the null hypothesis $H_0: \mu_1-\mu_2=0$.
For the treatment group, we have
\begin{align*}
\sqrt{m}\left(\frac{1}{m}\sum_{i=1}^{m}X_i-\mu_1\right) = \frac{1}{\sqrt{m}}\sum_{i=1}^m (X_i-\mu_1) +o_p(1)
\end{align*}
so that the influence function is $\varphi_1(x)=x-\mu_1$ and the asymptotic variance is $\tau_1^2=\sigma^2$.
Similarly, for the control group we have $\varphi_2(x)=x-\mu_2$ and $\tau_2^2=\sigma^2$.
To achieve a one-sided type I error rate $\alpha$ under the null hypothesis and power $1-\beta$ to detect $\delta_1$ under the alternative hypothesis, the required sample size for the control group is
\begin{align*}
n=\frac{(z_{1-\alpha}+z_{1-\beta})^2}{\delta^2_1}\left(\frac{\sigma^2}{r}+\sigma^2\right).
\end{align*}

For blinded sample size re-estimation, we need to determine the functions $g_1$ and $g_2$ in (\ref{eq-g1g2}).
Define the overall mean by $\bar{\mu}=(r\mu_1+\mu_2)/(r+1)$ and impose the constraint $\mu_1-\mu_2=\delta_1$.
Solving these two equations yields
\begin{align*}
  \mu_1
    = g_1(\bar{\mu},\delta_1)
    = \bar{\mu} + \frac{\delta_1}{r+1},
  \qquad
  \mu_2
    = g_2(\bar{\mu},\delta_1)
    = \bar{\mu} - \frac{r\,\delta_1}{r+1}.
\end{align*}
In the blinded setting, the nuisance parameters are $\bar{\mu}$ and $\sigma^2$.
This result is consistent with previous findings reported in \cite{GouldShih:1992},
\cite{KieserFriede:2003}, \cite{FriedeKieser:2006}, \cite[Chapter 11]{Proschanetal:2006}, and \cite{MuetzeFriede:2021}.

\end{exam}

\begin{exam}\label{example-2}
(Binary distribution)

Consider $X_i \sim \mathrm{Bernoulli}(p_1)$ and $Y_j \sim \mathrm{Bernoulli}(p_2)$.
We are interested in testing the null hypothesis $H_0: p_1-p_2=0$. For the treatment group, we have
\begin{align*}
\sqrt{m}\left(\frac{1}{m}\sum_{i=1}^{m}X_i-p_1\right) = \frac{1}{\sqrt{m}}\sum_{i=1}^m (X_i-p_1) +o_p(1)
\end{align*}
so that the influence function is $\varphi_1(x)=x-p_1$ and the asymptotic variance is $\tau_1^2=p_1(1-p_1)$.
Similarly, for the control group we have $\varphi_2(x)=x-p_2$ and $\tau_2^2=p_2(1-p_2)$.
To achieve a one-sided type I error rate $\alpha$ under the null hypothesis and power $1-\beta$ to detect $\delta_1$ under the alternative hypothesis, the required sample size for the control group is
\begin{align*}
n=\frac{(z_{1-\alpha}+z_{1-\beta})^2}{\delta^2_1}\left(\frac{p_1(1-p_1)}{r}+p_2(1-p_2)\right).
\end{align*}

For blinded sample size re-estimation, we need to determine the functions $g_1$ and $g_2$ in (\ref{eq-g1g2}).
Define the overall response rate by $\bar{p}=(rp_1+p_2)/(r+1)$ and impose the constraint $p_1-p_2=\delta_1$.
Solving these two equations yields
\begin{align*}
  p_1
    = g_1(\bar{p},\delta_1)
    = \bar{p} + \frac{\delta_1}{r+1},
  \qquad
  p_2
    = g_2(\bar{p},\delta_1)
    = \bar{p} - \frac{r\,\delta_1}{r+1}.
\end{align*}
The above representation is valid provided that the resulting $p_1$ and $p_2$ remain in $[0,1]$.
In the blinded setting, the nuisance parameter is $\bar{p}$.
This result is consistent with previous findings reported in \cite[Section 14.2]{JennisonTurnbull:1999}, \cite{FriedeKieser:2004}, \cite[Section 11.3]{Proschanetal:2006}, and \cite{Friedeetal:2007}.

We can consider several hypotheses. 
For example, we consider testing the null hypothesis $H_0: \log(p_1/(1-p_1))-\log(p_2/(1-p_2))=0$.
Let
\begin{align*}
\widehat{p}_1 =  \frac{1}{m}\sum_{i=1}^m X_i, \qquad \widehat{p}_2 =  \frac{1}{n}\sum_{j=1}^n Y_j.
\end{align*}
For the treatment group, we have
\begin{align*}
\sqrt{m}\left(\log\frac{\widehat p_1}{1-\widehat p_1}-\log\frac{p_1}{1-p_1}\right) = \frac{1}{\sqrt{m}}\sum_{i=1}^m\frac{X_i-p_1}{p_1(1-p_1)}+o_p(1)
\end{align*}
so that the influence function is $\varphi_1(x)=(x-p_1)/(p_1(1-p_1))$ and the asymptotic variance is $\tau_1^2=\mathbb{E}((X_i-p_1)^2/(p_1^2(1-p_1)^2))=p_1(1-p_1)/(p_1^2(1-p_1)^2)=1/(p_1(1-p_1))$.
Similarly, for the control group we have $\varphi_2(x)=(x-p_2)/(p_2(1-p_2))$ and $\tau_2^2=\mathbb{E}((Y_j-p_2)^2/(p_2^2(1-p_2)^2))=p_2(1-p_2)/(p_2^2(1-p_2)^2)=1/(p_2(1-p_2))$.
To achieve a one-sided type I error rate $\alpha$ under the null hypothesis and power $1-\beta$ to detect $\delta_1$ under the alternative hypothesis, the required sample size for the control group is
\begin{align*}
n=\frac{(z_{1-\alpha}+z_{1-\beta})^2}{\delta^2_1}\left(\frac{1}{p_1(1-p_1)r}+\frac{1}{p_2(1-p_2)}\right),
\end{align*}
which is consistent with \cite[Section 4.6]{Chowetal:2017}.

For blinded sample size re-estimation, we need to determine the functions $g_1$ and $g_2$ in (\ref{eq-g1g2}).
Define the overall response rate by $\bar{p}=(rp_1+p_2)/(r+1)$ and impose the constraint $\log\frac{p_1}{1-p_1}-\log\frac{p_2}{1-p_2}=\delta_1$.
Solving these two equations yields
\begin{align*}
  p_1
    &= g_1(\bar{p},\delta_1)
    = \frac{A p_2}{1+(A-1)p_2},
  \\
  p_2
    &= g_2(\bar{p},\delta_1)
    = \frac{-(rA+1-s(A-1))+\sqrt{(rA+1-s(A-1))^2+4(A-1)s}}{2(A-1)},
\end{align*}
where $A=e^{\delta_1}$ and $s=(r+1)\bar p$.
In the blinded setting, the nuisance parameter is $\bar{p}$.

We can also consider testing the null hypothesis $H_0: \log p_1-\log p_2=0$.
For the treatment group, we have
\begin{align*}
\sqrt{m}\left(\log\left(\frac{1}{m}\sum_{i=1}^{m}X_i \right)-\log p_1\right) = \frac{1}{\sqrt{m}}\sum_{i=1}^m\frac{X_i-p_1}{p_1}+o_p(1)
\end{align*}
so that the influence function is $\varphi_1(x)=(x-p_1)/p_1$ and the asymptotic variance is $\tau_1^2=\mathbb{E}((X_i-p_1)^2/p_1^2)=p_1(1-p_1)/p_1^2=(1-p_1)/p_1$.
Similarly, for the control group we have $\varphi_2(x)=(x-p_2)/p_2$ and $\tau_2^2=\mathbb{E}((Y_j-p_2)^2/p_2^2)=p_2(1-p_2)/p_2^2=(1-p_2)/p_2$.
To achieve a one-sided type I error rate $\alpha$ under the null hypothesis and power $1-\beta$ to detect $\delta_1$ under the alternative hypothesis, the required sample size for the control group is
\begin{align*}
n=\frac{(z_{1-\alpha}+z_{1-\beta})^2}{\delta^2_1}\left(\frac{1-p_1}{rp_1}+\frac{1-p_2}{p_2}\right),
\end{align*}
which is consistent with \cite[15.6.1]{Chowetal:2017} and \cite{Blackwelder:1993}. Similarly, we can define the nuisance parameter $\bar{p}$ in the blinded setting.


For blinded sample size re-estimation, we need to determine the functions $g_1$ and $g_2$ in (\ref{eq-g1g2}).
Define the overall response rate by $\bar{p}=(rp_1+p_2)/(r+1)$ and impose the constraint $\log p_1-\log p_2=\delta_1$.
Solving these two equations yields
\begin{align*}
  p_1
    = g_1(\bar{p},\delta_1)
    = \frac{e^{\delta_1}(r+1)\bar p}{r e^{\delta_1}+1},
  \qquad
  p_2
    = g_2(\bar{p},\delta_1)
    = \frac{(r+1)\bar p}{r e^{\delta_1}+1}.
\end{align*}
In the blinded setting, the nuisance parameter is $\bar{p}$.

\end{exam}

\begin{exam}\label{example-3}
(Negative binomial distribution)

Consider $X_{i}$ and $Y_{j}$ following negative binomial distribution
\begin{align*}
X_{i} \sim \mathrm{NB}(\lambda_1 t_{1i}, \ \phi), \qquad Y_{j} \sim \mathrm{NB}(\lambda_2 t_{2j}, \ \phi),
\end{align*}
where $\lambda_1$ and $\lambda_2$ denote the rate parameters, $t_{1i}$ and $t_{2j}$ are the exposure times, and $\phi$ is the dispersion parameter.
A natural estimator of $\lambda_1$ and $\lambda_2$ is given by
\begin{align*}
\widehat{\lambda}_1=\frac{\sum_{i=1}^{m_1}X_{i}}{\sum_{i=1}^{m_1}t_{1i}}, \qquad \widehat{\lambda}_2=\frac{\sum_{j=1}^{n_1}Y_{j}}{\sum_{j=1}^{n_1}t_{2j}},
\end{align*}

Applying the delta method yields
\begin{align}\label{eq-333}
\log\widehat{\lambda}_1-\log\lambda_1
= \frac{1}{\lambda_1 \sum_{i=1}^{m_1}t_{1i}} \sum_{i=1}^{m_1}(X_{i}-\lambda_1 t_{1i}) + o_p(m_1^{-1/2}),
\end{align}
and
\begin{align}\label{eq-334}
\log\widehat{\lambda}_2-\log\lambda_2
= \frac{1}{\lambda_2 \sum_{j=1}^{n_1}t_{2j}} \sum_{j=1}^{n_1}(Y_{j}-\lambda_2 t_{2j}) + o_p(n_1^{-1/2}),
\end{align}

For the negative binomial model, we have
\begin{align*}
\mathrm{Var}(X_{i})=\lambda_1 t_{1i}+\phi(\lambda_1 t_{1i})^2, \quad \mathrm{Var}(Y_{j})=\lambda_2 t_{2j}+\phi(\lambda_2 t_{2j})^2.
\end{align*}
Using (\ref{eq-333}) and (\ref{eq-334}), we obtain
\begin{align*}
\mathrm{Var}(\log\widehat{\lambda}_1)
&\approx
\frac{1}{(\lambda_1 \sum_{i=1}^{m_1}t_{1i})^2}
\sum_{i=1}^{m_1}
(\lambda_1 t_{1i} + \phi(\lambda_1 t_{1i})^2),\\
\mathrm{Var}(\log\widehat{\lambda}_2)
&\approx
\frac{1}{(\lambda_2 \sum_{j=1}^{n_1}t_{2j})^2}
\sum_{j=1}^{n_1}
(\lambda_2 t_{2j} + \phi(\lambda_2 t_{2j})^2).
\end{align*}
Since $\tau_1^2$ can be interpreted as the asymptotic variance of $\sqrt{m_1}(\log\widehat{\lambda}_1-\log\lambda_1)$, it follows that
\begin{align}\label{bn-tau}
\tau_1^2
=
m_1
\frac{1}{(\lambda_1 \sum_{i=1}^{m_1}t_{1i})^2}
\sum_{i=1}^{m_1}
(\lambda_1 t_{1i} + \phi(\lambda_1 t_{1i})^2)
=
\frac{\left(\frac{\sum_{i=1}^{m_1}t_{1i}}{m_1}\right)\frac{1}{\lambda_1}+\left(\frac{\sum_{i=1}^{m_1}t_{1i}^2}{m_1}\right)\phi}{\left(\frac{\sum_{i=1}^{m_1}t_{1i}}{m_1}\right)^2}.
\end{align}
Similarly,
\begin{align}\label{bn-tau-2}
\tau_2^2
=
n_1
\frac{1}{(\lambda_2 \sum_{j=1}^{n_1}t_{2j})^2}
\sum_{j=1}^{n_1}
(\lambda_2 t_{2j} + \phi(\lambda_2 t_{2j})^2)
=
\frac{\left(\frac{\sum_{j=1}^{n_1}t_{2j}}{n_1}\right)\frac{1}{\lambda_2}+\left(\frac{\sum_{j=1}^{n_1}t_{2j}^2}{n_1}\right)\phi}{\left(\frac{\sum_{j=1}^{n_1}t_{2j}}{n_1}\right)^2}.
\end{align}

We are interested in testing the null hypothesis $H_0: \log\lambda_1-\log\lambda_2=0$. To achieve a one-sided type I error rate $\alpha$ under the null hypothesis and power $1-\beta$ to detect $\delta_1$ under the alternative hypothesis, the required sample size for the control group is
\begin{align}\label{bn-n}
n=\frac{(z_{1-\alpha}+z_{1-\beta})^2}{\delta^2_1}\left(\frac{\tau_1^2}{r}+\tau_2^2\right),
\end{align}
where $\tau_1^2$ and $\tau_2^2$ are defined in (\ref{bn-tau}) and (\ref{bn-tau-2}), respectively.

For blinded sample size re-estimation, we need to determine the functions $g_1$ and $g_2$ in (\ref{eq-g1g2}). Define the overall event rate by $\bar{\lambda}=(r\lambda_1+\lambda_2)/(r+1)$ and impose the constraint $\log\lambda_1-\log\lambda_2=\delta_1$. Solving these two equations yields
\begin{align*}
  \lambda_1
    = g_1(\bar{\lambda},\delta_1)
    = \frac{(r+1)\bar{\lambda}e^{\delta_1}}{re^{\delta_1}+1} ,
  \qquad
  \lambda_2
    = g_2(\bar{\lambda},\delta_1)
    = \frac{(r+1)\bar{\lambda}}{re^{\delta_1}+1} .
\end{align*}
In the blinded setting, the nuisance parameters are $\bar{\lambda}$ and $\phi$.

Equation (\ref{bn-n}) provides a general formulation.
If we assume equal exposure times, $t_{1i}=t_{2j}=t$, then (\ref{bn-tau}) simplifies to
\begin{align*}
\tau_1^2 = \frac{1}{t\lambda_1}+\phi,\qquad
\tau_2^2 = \frac{1}{t\lambda_2}+\phi.
\end{align*}
Substituting these expressions into (\ref{bn-n}) yields a result identical to that reported in \cite{Keeneetal:2007}, \cite{MuetzeFriede:2021}, and \cite{FriedeSchmidli:2010b}.
Note that the Poisson distribution is a special case of the negative binomial distribution obtained by setting $\phi=0$, and the resulting formula is consistent with \cite{FriedeSchmidli:2010} and \cite{Schneideretal:2013b}.

\end{exam}

From Example \ref{example-1} to Example \ref{example-3}, the proposed sample size re-estimation framework recovers and unifies several existing approaches discussed above.
We next demonstrate how the proposed framework can also be used to derive new procedures.

For continuous outcomes that are positive and skewed, it is common to assume distributions such as the Gamma, inverse Gaussian, or lognormal distribution, depending on the scientific context.
Within this general framework, blinded sample size re-estimation procedures can be developed for continuous outcomes following Gamma, inverse Gaussian, or lognormal distributions, which have received relatively limited attention in the existing literature.

\begin{exam}\label{example-4}
(Gamma distribution)

Consider $X_i \sim \mathrm{Gamma}(\mu_1,\phi)$ and $Y_j \sim \mathrm{Gamma}(\mu_2,\phi)$, where $\mathbb{E}(X_i)=\mu_1$, $\mathbb{E}(Y_j)=\mu_2$, $\mathrm{Var}(X_i)=\phi\mu_1^2$, and $\mathrm{Var}(Y_j)=\phi\mu_2^2$.
We are interested in testing the null hypothesis $H_0: \log\mu_1-\log\mu_2=0$. For the treatment group, we have
\begin{align*}
\sqrt{m}\left(\log\!\left(\frac{1}{m}\sum_{i=1}^{m} X_i\right)-\log \mu_1\right) = \frac{1}{\sqrt{m}}\sum_{i=1}^m\frac{X_i-\mu_1}{\mu_1}+o_p(1)
\end{align*}
so that the influence function is $\varphi_1(x)=(x-\mu_1)/\mu_1$ and the asymptotic variance is $\tau_1^2=\mathbb{E}((X_i-\mu_1)^2/\mu_1^2)=\mathrm{Var}(X_i)/\mu_1^2=\phi$.
Similarly, for the control group we have $\varphi_2(x)=(x-\mu_2)/\mu_2$ and $\tau_2^2=\mathbb{E}((Y_j-\mu_2)^2/\mu_2^2)=\mathrm{Var}(Y_j)/\mu_2^2=\phi$.
To achieve a one-sided type I error rate $\alpha$ under the null hypothesis and power $1-\beta$ to detect $\delta_1$ under the alternative hypothesis, the required sample size for the control group is
\begin{align*}
n=\frac{(z_{1-\alpha}+z_{1-\beta})^2}{\delta^2_1}\left(\frac{\phi}{r}+\phi\right),
\end{align*}
which is consistent with \cite{CundillAlexander:2015}.

For blinded sample size re-estimation, we need to determine the functions $g_1$ and $g_2$ in (\ref{eq-g1g2}). Define the overall mean by $\bar{\mu}=(r\mu_1+\mu_2)/(r+1)$ and impose the constraint $\log\mu_1-\log\mu_2=\delta_1$. Solving these two equations yields
\begin{align*}
  \mu_1
    = g_1(\bar{\mu},\delta_1)
    = \frac{(r+1)\bar{\mu}}{re^{\delta_1}+1}e^{\delta_1},
  \qquad
  \mu_2
    = g_2(\bar{\mu},\delta_1)
    = \frac{(r+1)\bar{\mu}}{re^{\delta_1}+1}.
\end{align*}
In the blinded setting, the nuisance parameters are $\bar{\mu}$ and $\phi$.

\end{exam}

\begin{exam}\label{example-5}
(Inverse Gaussian distribution)

Consider $X_i \sim \mathrm{IG}(\mu_1,\lambda)$ and $Y_j \sim \mathrm{IG}(\mu_2,\lambda)$, where $\mathbb{E}(X_i)=\mu_1$, $\mathbb{E}(Y_j)=\mu_2$, $\mathrm{Var}(X_i)=\mu_1^3/\lambda$, and $\mathrm{Var}(Y_j)=\mu_2^3/\lambda$. Here, $\lambda$ denotes the common shape parameter.
We are interested in testing the null hypothesis $H_0: \log\mu_1-\log\mu_2=0$. For the treatment group, we have
\begin{align*}
\sqrt{m}\left(\log\!\left(\frac{1}{m}\sum_{i=1}^{m} X_i\right)-\log \mu_1\right) = \frac{1}{\sqrt{m}}\sum_{i=1}^m\frac{X_i-\mu_1}{\mu_1}+o_p(1)
\end{align*}
so that the influence function is $\varphi_1=(x-\mu_1)/\mu_1$ and the asymptotic variance is $\tau_1^2=\mathbb{E}((X_i-\mu_1)^2/\mu_1^2)=\mathrm{Var}(X_i)/\mu_1^2=\mu_1/\lambda$.
Similarly, for the control group we have $\varphi_2=(x-\mu_2)/\mu_2$ and $\tau_2^2=\mathbb{E}((Y_j-\mu_2)^2/\mu_2^2)=\mathrm{Var}(Y_j)/\mu_2^2=\mu_2/\lambda$.
To achieve a one-sided type I error rate $\alpha$ under the null hypothesis and power $1-\beta$ to detect $\delta_1$ under the alternative hypothesis, the required sample size for the control group is
\begin{align*}
n=\frac{(z_{1-\alpha}+z_{1-\beta})^2}{\delta^2_1}\left(\frac{\mu_1}{r\lambda}+\frac{\mu_2}{\lambda}\right).
\end{align*}

For blinded sample size re-estimation, we need to determine the functions $g_1$ and $g_2$ in (\ref{eq-g1g2}). Define the overall mean by $\bar{\mu}=(r\mu_1+\mu_2)/(r+1)$ and impose the constraint $\log\mu_1-\log\mu_2=\delta_1$. Solving these two equations yields
\begin{align*}
  \mu_1
    = g_1(\bar{\mu},\delta_1)
    = \frac{(r+1)\bar{\mu}}{re^{\delta_1}+1}e^{\delta_1},
  \qquad
  \mu_2
    = g_2(\bar{\mu},\delta_1)
    = \frac{(r+1)\bar{\mu}}{re^{\delta_1}+1}.
\end{align*}
In the blinded setting, the nuisance parameters are $\bar{\mu}$ and $\lambda$.

\end{exam}

\begin{exam}\label{example-6}
(Lognormal distribution)

Consider $X_i \sim \mathrm{Lognormal}(\mu_1,\sigma^2)$ and
$Y_j \sim \mathrm{Lognormal}(\mu_2,\sigma^2)$, where $\mathbb{E}(X_i)=\exp(\mu_1+\sigma^2/2)$ and $\mathbb{E}(Y_j)=\exp(\mu_2+\sigma^2/2)$.
Here, $\mu_1$ and $\mu_2$ denote the location parameters of the lognormal distributions, and $\sigma^2$ is the common variance parameter on the logarithmic scale.
Equivalently, $\log X_i \sim \mathcal{N}(\mu_1,\sigma^2)$ and $\log Y_j \sim \mathcal{N}(\mu_2,\sigma^2)$.
Note that $\sigma^2$ is the variance of $\log X_i$ rather than the variance of $X_i$.

We are interested in testing the null hypothesis $H_0: \mu_1-\mu_2=0$. For the treatment group, we have
\begin{align*}
\sqrt{m}\left(\frac{1}{m}\sum_{i=1}^{m}\log X_i-\mu_1\right) = \frac{1}{\sqrt{m}}\sum_{i=1}^m (\log X_i-\mu_1) +o_p(1)
\end{align*}
so that the influence function is $\varphi_1(x)=\log x-\mu_1$ and the asymptotic variance is $\tau_1^2=\mathbb{E}((\log X_i-\mu_1)^2)=\sigma^2$.
Similarly, for the control group we have $\varphi_2(x)=\log x-\mu_2$ and $\tau_2^2=\mathbb{E}((\log Y_j-\mu_2)^2)=\sigma^2$.
To achieve a one-sided type I error rate $\alpha$ under the null hypothesis and power $1-\beta$ to detect $\delta_1$ under the alternative hypothesis, the required sample size for the control group is
\begin{align*}
n=\frac{(z_{1-\alpha}+z_{1-\beta})^2}{\delta^2_1}\left(\frac{\sigma^2}{r}+\sigma^2\right),
\end{align*}
which is consistent with \cite{OKeeffeetal:2017} and \cite{WolfeCarlin:1999}.

For blinded sample size re-estimation, we need to determine the functions $g_1$ and $g_2$ in (\ref{eq-g1g2}). Define the pooled quantity on the exponential scale by $\exp(\bar{\mu})=(r\exp(\mu_1)+\exp(\mu_2))/(r+1)$ and impose the constraint $\mu_1-\mu_2=\delta_1$. Solving these two equations yields
\begin{align*}
  \mu_1
    = g_1(\bar{\mu},\delta_1)
    = \log\frac{(r+1)e^{\bar{\mu}}}{r e^{\delta_1} +1} + \delta_1,
  \qquad
  \mu_2
    = g_2(\bar{\mu},\delta_1)
    = \log\frac{(r+1)e^{\bar{\mu}}}{r e^{\delta_1} +1}.
\end{align*}
In the blinded setting, the nuisance parameters are $\bar{\mu}$ and $\sigma^2$.

\end{exam}

\section{About Huber's Proposal 2 Estimator}

Let $y_1,\ldots,y_n$ denote independent observations. Huber's Proposal 2 estimator jointly estimates a robust location parameter $\mu$ and a robust scale parameter $\sigma$ by solving estimating equations based on the Huber score function. For a tuning constant $k>0$, define
\begin{align*}
\psi_k(e)=
\begin{cases}
e, & |e|\leq k,\\
k\,\mathrm{sign}(e), & |e|>k,
\end{cases}
\end{align*}
and $\chi_k(e)=\psi_k(e)^2=\min(e^2,k^2)$.

The standardized residuals are
\begin{align*}
e_i=\frac{y_i-\mu}{\sigma}, \qquad i=1,\ldots,n.
\end{align*}

Huber's Proposal 2 estimator $(\hat{\mu},\hat{\sigma})$ is defined as the solution to
\begin{align*}
\sum_{i=1}^n \psi_k\left(\frac{y_i-\mu}{\sigma}\right)=0,
\end{align*}
and
\begin{align*}
\frac{1}{n}\sum_{i=1}^n
\chi_k\left(\frac{y_i-\mu}{\sigma}\right)=b_k,
\end{align*}
where $b_k$ is a normalizing constant chosen to make the scale estimator consistent under normality. If $Z\sim \mathcal{N}(0,1)$, then
\begin{align*}
b_k=\mathbb{E}(\chi_k(Z))
=\mathbb{E}(\min(Z^2,k^2))
=
2\Phi(k)-1-2k\phi(k)+2k^2(1-\Phi(k)),
\end{align*}
where $\phi(\cdot)$ and $\Phi(\cdot)$ are the standard normal density and distribution function, respectively.

\section{Additional Simulation}

\subsection{Type I error and power}

Tables~\ref{tab:combined_type1_sample_h0_sigma12_v2} and~\ref{tab:combined_power_sample_h1_sigma12} present the empirical type I error rates and power, respectively, for the unequal-variance setting with $\sigma_1=1$ and $\sigma_2=2$.

\begin{table}[htbp]
\centering
\footnotesize
\caption{Simulation under the null hypothesis ($\mu_1=\mu_2=0$, $\sigma_1=1$, $\sigma_2=2$)}
\label{tab:combined_type1_sample_h0_sigma12_v2}
\resizebox{\textwidth}{!}{
\begin{tabular}{llccccccccc}
\toprule
& & \multicolumn{3}{c}{$n_1=5$}
& \multicolumn{3}{c}{$n_1=10$}
& \multicolumn{3}{c}{$n_1=50$} \\
\cmidrule(lr){3-5} \cmidrule(lr){6-8} \cmidrule(lr){9-11}
$\epsilon$ & Method
& Type I & Mean($N$) & SD($N$)
& Type I & Mean($N$) & SD($N$)
& Type I & Mean($N$) & SD($N$) \\
\midrule

\multirow{6}{*}{$0$}
& OS
& 0.02509 & 157.66 & 84.66
& 0.02615 & 157.54 & 59.11
& 0.02560 & 157.41 & 26.00 \\

& UN
& 0.02600 & 157.70 & 91.15
& 0.02526 & 157.56 & 61.10
& 0.02658 & 157.41 & 26.16 \\

& Robust ($k=1.5$)
& 0.02546 & 155.46 & 89.30
& 0.02530 & 141.52 & 57.32
& 0.02614 & 131.54 & 23.40 \\

& Robust ($k=1.345$)
& 0.02572 & 155.30 & 91.32
& 0.02587 & 139.47 & 57.97
& 0.02473 & 128.50 & 23.52 \\

& Robust ($k=1.2$)
& 0.02520 & 155.45 & 94.38
& 0.02489 & 138.07 & 59.36
& 0.02468 & 125.95 & 23.95 \\

\midrule

\multirow{6}{*}{$0.1$}
& OS
& 0.02463 & 203.72 & 142.64
& 0.02525 & 204.92 & 103.20
& 0.02508 & 204.53 & 45.45 \\

& UN
& 0.02542 & 203.66 & 146.33
& 0.02552 & 204.86 & 104.31
& 0.02555 & 204.54 & 45.57 \\

& Robust ($k=1.5$)
& 0.02442 & 185.31 & 119.34
& 0.02508 & 166.45 & 73.21
& 0.02433 & 151.87 & 29.11 \\

& Robust ($k=1.345$)
& 0.02430 & 183.82 & 119.77
& 0.02576 & 162.90 & 72.95
& 0.02355 & 147.38 & 28.84 \\

& Robust ($k=1.2$)
& 0.02521 & 182.76 & 121.57
& 0.02476 & 160.19 & 73.64
& 0.02518 & 143.58 & 28.94 \\

\midrule

\multirow{6}{*}{$0.2$}
& OS
& 0.02431 & 250.13 & 179.26
& 0.02465 & 252.38 & 131.70
& 0.02570 & 251.61 & 58.38 \\

& UN
& 0.02553 & 249.76 & 182.24
& 0.02570 & 252.36 & 132.96
& 0.02587 & 251.60 & 58.48 \\

& Robust ($k=1.5$)
& 0.02408 & 220.49 & 150.13
& 0.02487 & 194.56 & 91.24
& 0.02577 & 175.89 & 35.74 \\

& Robust ($k=1.345$)
& 0.02488 & 217.72 & 149.87
& 0.02494 & 189.61 & 90.26
& 0.02463 & 169.75 & 35.01 \\

& Robust ($k=1.2$)
& 0.02419 & 215.66 & 151.32
& 0.02507 & 185.64 & 90.29
& 0.02473 & 164.48 & 34.71 \\

\midrule

\multirow{6}{*}{$0.3$}
& OS
& 0.02479 & 295.66 & 204.84
& 0.02489 & 298.97 & 153.36
& 0.02455 & 299.06 & 67.98 \\

& UN
& 0.02549 & 295.06 & 207.77
& 0.02543 & 298.96 & 154.90
& 0.02590 & 299.05 & 68.13 \\

& Robust ($k=1.5$)
& 0.02484 & 258.09 & 176.13
& 0.02576 & 227.17 & 109.73
& 0.02480 & 204.22 & 42.76 \\

& Robust ($k=1.345$)
& 0.02448 & 254.11 & 175.34
& 0.02466 & 220.65 & 108.11
& 0.02539 & 196.28 & 41.69 \\

& Robust ($k=1.2$)
& 0.02548 & 250.86 & 176.59
& 0.02563 & 215.44 & 107.91
& 0.02514 & 189.38 & 41.09 \\

\bottomrule
\end{tabular}
}
\end{table}

\begin{table}[htbp]
\centering
\footnotesize
\caption{Simulation under the alternative hypothesis ($\mu_1=0.5$, $\mu_2=0$, $\sigma_1=1$, $\sigma_2=2$)}
\label{tab:combined_power_sample_h1_sigma12}
\resizebox{\textwidth}{!}{
\begin{tabular}{llccccccccc}
\toprule
& & \multicolumn{3}{c}{$n_1=5$}
& \multicolumn{3}{c}{$n_1=10$}
& \multicolumn{3}{c}{$n_1=50$} \\
\cmidrule(lr){3-5} \cmidrule(lr){6-8} \cmidrule(lr){9-11}
$\epsilon$ & Method
& Power & Mean($N$) & SD($N$)
& Power & Mean($N$) & SD($N$)
& Power & Mean($N$) & SD($N$) \\
\midrule

\multirow{6}{*}{$0$}
& OS
& 0.72686 & 161.30 & 86.76
& 0.76625 & 161.61 & 60.18
& 0.79788 & 161.34 & 26.32 \\

& UN
& 0.71457 & 157.16 & 91.63
& 0.76053 & 157.45 & 61.08
& 0.79552 & 157.38 & 26.05 \\

& Robust ($k=1.5$)
& 0.71428 & 160.10 & 92.90
& 0.71690 & 146.12 & 59.09
& 0.72422 & 135.88 & 24.32 \\

& Robust ($k=1.345$)
& 0.71116 & 160.00 & 95.29
& 0.71149 & 144.10 & 59.97
& 0.71307 & 132.76 & 24.51 \\

& Robust ($k=1.2$)
& 0.70477 & 160.25 & 98.61
& 0.70521 & 142.76 & 61.60
& 0.70410 & 130.10 & 24.99 \\

\midrule

\multirow{6}{*}{$0.1$}
& OS
& 0.70091 & 208.50 & 144.65
& 0.74736 & 208.67 & 104.22
& 0.79279 & 208.63 & 46.22 \\

& UN
& 0.69335 & 204.29 & 147.70
& 0.74184 & 204.49 & 104.65
& 0.79237 & 204.67 & 45.95 \\

& Robust ($k=1.5$)
& 0.67852 & 191.15 & 122.10
& 0.67560 & 170.79 & 75.33
& 0.67491 & 156.54 & 30.02 \\

& Robust ($k=1.345$)
& 0.67511 & 189.84 & 122.97
& 0.66780 & 167.40 & 75.32
& 0.66391 & 152.05 & 29.81 \\

& Robust ($k=1.2$)
& 0.66911 & 189.08 & 125.50
& 0.66422 & 164.85 & 76.19
& 0.65298 & 148.18 & 29.93 \\

\midrule

\multirow{6}{*}{$0.2$}
& OS
& 0.68970 & 253.86 & 179.36
& 0.74037 & 255.45 & 131.28
& 0.78984 & 255.48 & 58.68 \\

& UN
& 0.68522 & 249.54 & 182.03
& 0.73612 & 251.22 & 131.60
& 0.79010 & 251.55 & 58.48 \\

& Robust ($k=1.5$)
& 0.66117 & 224.95 & 150.70
& 0.65694 & 199.77 & 92.74
& 0.65219 & 180.92 & 36.51 \\

& Robust ($k=1.345$)
& 0.65309 & 222.51 & 150.89
& 0.64627 & 195.02 & 92.10
& 0.63774 & 174.82 & 35.93 \\

& Robust ($k=1.2$)
& 0.64441 & 220.80 & 152.80
& 0.63743 & 191.31 & 92.65
& 0.62213 & 169.56 & 35.78 \\

\midrule

\multirow{6}{*}{$0.3$}
& OS
& 0.68956 & 299.72 & 206.19
& 0.73768 & 302.92 & 153.05
& 0.78904 & 302.80 & 68.30 \\

& UN
& 0.68267 & 294.69 & 208.35
& 0.73340 & 298.82 & 153.80
& 0.78784 & 298.85 & 68.15 \\

& Robust ($k=1.5$)
& 0.65176 & 263.70 & 179.33
& 0.64709 & 232.82 & 111.72
& 0.63807 & 209.34 & 43.85 \\

& Robust ($k=1.345$)
& 0.64416 & 260.07 & 179.15
& 0.63749 & 226.65 & 110.53
& 0.62249 & 201.55 & 42.92 \\

& Robust ($k=1.2$)
& 0.63568 & 257.10 & 180.59
& 0.62808 & 221.74 & 110.56
& 0.60934 & 194.78 & 42.48 \\

\bottomrule
\end{tabular}
}
\end{table}

Tables~\ref{tab:combined_type1_sample_unequal} and~\ref{tab:combined_results_unequal} present the empirical type I error rates and power, respectively, for the unequal-dispersion setting with $\phi_1=0.6$ and $\phi_2=2$.

\begin{table}[htbp]
\centering
\footnotesize
\caption{Simulation under the null hypothesis ($\lambda_1=\lambda_2=2$, $\phi_1=0.6$, $\phi_2=2$)}
\label{tab:combined_type1_sample_unequal}
\resizebox{\textwidth}{!}{
\begin{tabular}{llccccccccc}
\toprule
& & \multicolumn{3}{c}{$n_1=5$}
& \multicolumn{3}{c}{$n_1=10$}
& \multicolumn{3}{c}{$n_1=50$} \\
\cmidrule(lr){3-5} \cmidrule(lr){6-8} \cmidrule(lr){9-11}
$\epsilon$ & Method
& Type I & Mean($N$) & SD($N$)
& Type I & Mean($N$) & SD($N$)
& Type I & Mean($N$) & SD($N$) \\
\midrule

\multirow{6}{*}{$0$}
& OS
& 0.02467 & 321.93 & 184.53
& 0.02522 & 322.79 & 131.31
& 0.02506 & 323.18 & 55.24 \\

& UN
& 0.02559 & 317.11 & 197.14
& 0.02578 & 309.96 & 129.52
& 0.02489 & 318.25 & 54.81 \\

& Robust ($k=1.5$)
& 0.02417 & 299.39 & 165.99
& 0.02565 & 272.31 & 111.56
& 0.02576 & 247.35 & 31.99 \\

& Robust ($k=1.345$)
& 0.02483 & 309.05 & 164.90
& 0.02454 & 282.75 & 110.79
& 0.02481 & 258.01 & 32.97 \\

& Robust ($k=1.2$)
& 0.02529 & 325.12 & 164.46
& 0.02545 & 299.33 & 110.20
& 0.02454 & 275.07 & 34.87 \\

\midrule

\multirow{6}{*}{$0.1$}
& OS
& 0.02488 & 335.04 & 191.66
& 0.02527 & 337.51 & 137.72
& 0.02539 & 336.98 & 58.15 \\

& UN
& 0.02557 & 328.52 & 201.70
& 0.02547 & 323.92 & 135.42
& 0.02523 & 331.89 & 57.63 \\

& Robust ($k=1.5$)
& 0.02506 & 309.52 & 173.16
& 0.02546 & 282.84 & 119.73
& 0.02562 & 254.40 & 34.09 \\

& Robust ($k=1.345$)
& 0.02573 & 319.02 & 171.81
& 0.02550 & 293.27 & 118.54
& 0.02434 & 265.17 & 34.93 \\

& Robust ($k=1.2$)
& 0.02542 & 335.08 & 170.82
& 0.02602 & 309.95 & 117.25
& 0.02624 & 282.53 & 36.57 \\

\midrule

\multirow{6}{*}{$0.2$}
& OS
& 0.02518 & 346.79 & 198.03
& 0.02611 & 351.83 & 144.73
& 0.02522 & 350.46 & 61.02 \\

& UN
& 0.02551 & 339.57 & 207.49
& 0.02606 & 337.37 & 141.64
& 0.02539 & 345.24 & 60.43 \\

& Robust ($k=1.5$)
& 0.02654 & 318.83 & 180.11
& 0.02582 & 293.69 & 129.53
& 0.02536 & 261.36 & 36.76 \\

& Robust ($k=1.345$)
& 0.02648 & 328.29 & 178.52
& 0.02496 & 304.04 & 127.97
& 0.02555 & 272.27 & 37.42 \\

& Robust ($k=1.2$)
& 0.02567 & 344.41 & 177.14
& 0.02611 & 320.63 & 125.99
& 0.02574 & 289.92 & 38.78 \\

\midrule

\multirow{6}{*}{$0.3$}
& OS
& 0.02726 & 361.00 & 205.05
& 0.02603 & 365.80 & 150.85
& 0.02603 & 364.37 & 63.71 \\

& UN
& 0.02618 & 352.04 & 212.37
& 0.02484 & 350.63 & 147.13
& 0.02635 & 358.96 & 63.09 \\

& Robust ($k=1.5$)
& 0.02597 & 330.68 & 188.27
& 0.02617 & 304.41 & 138.57
& 0.02648 & 268.60 & 40.17 \\

& Robust ($k=1.345$)
& 0.02689 & 340.10 & 186.37
& 0.02548 & 314.67 & 136.60
& 0.02682 & 279.58 & 40.53 \\

& Robust ($k=1.2$)
& 0.02557 & 356.27 & 184.38
& 0.02625 & 331.20 & 133.98
& 0.02654 & 297.48 & 41.49 \\

\bottomrule
\end{tabular}
}
\end{table}

\begin{table}[htbp]
\centering
\footnotesize
\caption{Simulation under the alternative hypothesis ($\lambda_1=1.5$, $\lambda_2=2$, $\phi_1=0.6$, $\phi_2=2$)}
\label{tab:combined_results_unequal}
\resizebox{\textwidth}{!}{
\begin{tabular}{llccccccccc}
\toprule
& & \multicolumn{3}{c}{$n_1=5$}
& \multicolumn{3}{c}{$n_1=10$}
& \multicolumn{3}{c}{$n_1=50$} \\
\cmidrule(lr){3-5} \cmidrule(lr){6-8} \cmidrule(lr){9-11}
$\epsilon$ & Method
& Power & Mean($N$) & SD($N$)
& Power & Mean($N$) & SD($N$)
& Power & Mean($N$) & SD($N$) \\
\midrule

\multirow{6}{*}{$0$}
& OS
& 0.71664 & 356.39 & 203.28
& 0.74911 & 358.78 & 150.72
& 0.78365 & 358.85 & 64.01 \\

& UN
& 0.70526 & 351.67 & 212.11
& 0.73754 & 344.12 & 146.30
& 0.77901 & 350.88 & 62.08 \\

& Robust ($k=1.5$)
& 0.70022 & 331.70 & 188.83
& 0.69173 & 304.48 & 139.84
& 0.67545 & 267.82 & 44.09 \\

& Robust ($k=1.345$)
& 0.71673 & 341.11 & 186.74
& 0.71068 & 314.72 & 137.71
& 0.69395 & 279.16 & 44.24 \\

& Robust ($k=1.2$)
& 0.73660 & 356.74 & 184.36
& 0.73320 & 331.13 & 134.83
& 0.72305 & 297.35 & 44.64 \\

\midrule

\multirow{6}{*}{$0.1$}
& OS
& 0.71602 & 368.15 & 209.00
& 0.75132 & 371.77 & 156.08
& 0.78625 & 371.23 & 66.54 \\

& UN
& 0.70437 & 362.51 & 217.38
& 0.73960 & 356.40 & 151.67
& 0.78184 & 363.00 & 64.57 \\

& Robust ($k=1.5$)
& 0.69759 & 341.26 & 195.11
& 0.69099 & 315.43 & 147.87
& 0.67116 & 275.43 & 49.40 \\

& Robust ($k=1.345$)
& 0.70960 & 350.55 & 192.83
& 0.70683 & 325.50 & 145.41
& 0.68967 & 286.83 & 49.08 \\

& Robust ($k=1.2$)
& 0.73329 & 366.13 & 190.08
& 0.73021 & 341.71 & 141.96
& 0.71565 & 305.20 & 48.80 \\

\midrule

\multirow{6}{*}{$0.2$}
& OS
& 0.71453 & 380.25 & 214.80
& 0.75266 & 385.24 & 162.06
& 0.78949 & 383.79 & 69.09 \\

& UN
& 0.70510 & 373.25 & 221.36
& 0.73661 & 369.06 & 157.26
& 0.78513 & 375.33 & 67.16 \\

& Robust ($k=1.5$)
& 0.69539 & 351.92 & 202.27
& 0.69156 & 326.87 & 156.79
& 0.67029 & 283.84 & 56.16 \\

& Robust ($k=1.345$)
& 0.70772 & 361.14 & 199.77
& 0.70692 & 336.86 & 153.99
& 0.68713 & 295.24 & 55.31 \\

& Robust ($k=1.2$)
& 0.73025 & 376.64 & 196.58
& 0.72788 & 352.97 & 149.94
& 0.71308 & 313.66 & 54.19 \\

\midrule

\multirow{6}{*}{$0.3$}
& OS
& 0.71268 & 390.22 & 218.74
& 0.75216 & 398.24 & 167.61
& 0.78885 & 396.31 & 71.63 \\

& UN
& 0.70042 & 383.47 & 225.94
& 0.73827 & 381.54 & 162.73
& 0.78306 & 387.63 & 69.69 \\

& Robust ($k=1.5$)
& 0.69435 & 360.76 & 207.22
& 0.69305 & 338.55 & 165.42
& 0.66896 & 292.36 & 63.22 \\

& Robust ($k=1.345$)
& 0.70450 & 369.93 & 204.50
& 0.70256 & 348.36 & 162.32
& 0.68516 & 303.71 & 61.89 \\

& Robust ($k=1.2$)
& 0.72767 & 385.42 & 200.92
& 0.72686 & 364.27 & 157.74
& 0.71105 & 322.17 & 59.97 \\

\bottomrule
\end{tabular}
}
\end{table}

\subsection{Relative efficiency ratios}

The relative efficiency ratios, calculated relative to the OS method, for the simulation settings described above are presented in Tables~\ref{tab:er_vs_os_sigma11_h1}, \ref{tab:er_vs_os_sigma12_h1}, \ref{tab:er_vs_os_nb_equal}, and~\ref{tab:er_vs_os_nb_unequal}.

\begin{table}[htbp]
\centering
\footnotesize
\caption{Relative efficiency ratios (ER) of each method compared with OS under the alternative hypothesis ($\mu_1=0.5$, $\mu_2=0$, $\sigma_1=\sigma_2=1$) for the one-sided test ($\alpha=0.025$).}
\label{tab:er_vs_os_sigma11_h1}
\begin{tabular}{llccc}
\toprule
& & \multicolumn{3}{c}{$n_1$} \\
\cmidrule(lr){3-5}
$\epsilon$ & Method & 5 & 10 & 50 \\
\midrule
\multirow{6}{*}{$0$}
& OS & 100.00 & 100.00 & 100.00 \\
& UN & 104.84 & 104.81 & 104.73 \\
& Robust ($k=1.5$) & 95.64 & 97.37 & 99.35 \\
& Robust ($k=1.345$) & 93.29 & 96.40 & 99.43 \\
& Robust ($k=1.2$) & 91.44 & 95.53 & 99.05 \\
\midrule
\multirow{6}{*}{$0.1$}
& OS & 100.00 & 100.00 & 100.00 \\
& UN & 101.29 & 101.95 & 103.03 \\
& Robust ($k=1.5$) & 122.94 & 118.62 & 103.56 \\
& Robust ($k=1.345$) & 122.58 & 118.89 & 103.58 \\
& Robust ($k=1.2$) & 121.37 & 117.90 & 103.50 \\
\midrule
\multirow{6}{*}{$0.2$}
& OS & 100.00 & 100.00 & 100.00 \\
& UN & 100.81 & 101.77 & 102.24 \\
& Robust ($k=1.5$) & 128.21 & 124.93 & 105.58 \\
& Robust ($k=1.345$) & 130.41 & 124.47 & 105.90 \\
& Robust ($k=1.2$) & 129.72 & 124.29 & 105.80 \\
\midrule
\multirow{6}{*}{$0.3$}
& OS & 100.00 & 100.00 & 100.00 \\
& UN & 100.25 & 100.40 & 101.03 \\
& Robust ($k=1.5$) & 124.39 & 122.24 & 105.56 \\
& Robust ($k=1.345$) & 127.06 & 122.85 & 105.75 \\
& Robust ($k=1.2$) & 128.08 & 123.20 & 106.08 \\
\bottomrule
\end{tabular}
\end{table}

\begin{table}[htbp]
\centering
\footnotesize
\caption{Relative efficiency ratios (ER) of each method compared with OS under the alternative hypothesis ($\mu_1=0.5$, $\mu_2=0$, $\sigma_1=1$, $\sigma_2=2$) for the one-sided test ($\alpha=0.025$).}
\label{tab:er_vs_os_sigma12_h1}
\begin{tabular}{llccc}
\toprule
& & \multicolumn{3}{c}{$n_1$} \\
\cmidrule(lr){3-5}
$\epsilon$ & Method & 5 & 10 & 50 \\
\midrule
\multirow{6}{*}{$0$}
& OS & 100.00 & 100.00 & 100.00 \\
& UN & 99.73 & 101.23 & 101.90 \\
& Robust ($k=1.5$) & 97.83 & 98.37 & 99.32 \\
& Robust ($k=1.345$) & 97.18 & 98.50 & 99.04 \\
& Robust ($k=1.2$) & 95.60 & 97.98 & 98.98 \\
\midrule
\multirow{6}{*}{$0.1$}
& OS & 100.00 & 100.00 & 100.00 \\
& UN & 100.29 & 100.71 & 101.83 \\
& Robust ($k=1.5$) & 103.58 & 103.35 & 100.73 \\
& Robust ($k=1.345$) & 103.48 & 103.57 & 101.11 \\
& Robust ($k=1.2$) & 102.47 & 104.31 & 101.18 \\
\midrule
\multirow{6}{*}{$0.2$}
& OS & 100.00 & 100.00 & 100.00 \\
& UN & 100.69 & 100.67 & 101.63 \\
& Robust ($k=1.5$) & 105.68 & 105.36 & 102.05 \\
& Robust ($k=1.345$) & 104.87 & 105.31 & 102.15 \\
& Robust ($k=1.2$) & 103.59 & 105.18 & 101.59 \\
\midrule
\multirow{6}{*}{$0.3$}
& OS & 100.00 & 100.00 & 100.00 \\
& UN & 100.10 & 100.35 & 101.02 \\
& Robust ($k=1.5$) & 104.19 & 105.47 & 101.39 \\
& Robust ($k=1.345$) & 103.81 & 105.97 & 101.58 \\
& Robust ($k=1.2$) & 102.97 & 105.99 & 101.95 \\
\bottomrule
\end{tabular}
\end{table}

\begin{table}[htbp]
\centering
\footnotesize
\caption{Relative efficiency ratios (ER) of each method compared with OS under the alternative hypothesis ($\lambda_1=1.5$, $\lambda_2=2$, $\phi_1=\phi_2=0.6$) for the one-sided test ($\alpha=0.025$).}
\label{tab:er_vs_os_nb_equal}
\begin{tabular}{llccc}
\toprule
& & \multicolumn{3}{c}{$n_1$} \\
\cmidrule(lr){3-5}
$\epsilon$ & Method & 5 & 10 & 50 \\
\midrule
\multirow{6}{*}{$0$}
& OS & 100.00 & 100.00 & 100.00 \\
& UN & 99.45 & 101.44 & 100.96 \\
& Robust ($k=1.5$) & 103.89 & 103.90 & 101.63 \\
& Robust ($k=1.345$) & 103.57 & 103.60 & 100.97 \\
& Robust ($k=1.2$) & 102.99 & 102.89 & 101.25 \\
\midrule
\multirow{6}{*}{$0.1$}
& OS & 100.00 & 100.00 & 100.00 \\
& UN & 98.56 & 100.53 & 100.65 \\
& Robust ($k=1.5$) & 103.75 & 103.57 & 101.01 \\
& Robust ($k=1.345$) & 103.43 & 103.59 & 101.42 \\
& Robust ($k=1.2$) & 102.98 & 102.73 & 100.53 \\
\midrule
\multirow{6}{*}{$0.2$}
& OS & 100.00 & 100.00 & 100.00 \\
& UN & 98.53 & 100.88 & 100.78 \\
& Robust ($k=1.5$) & 103.59 & 103.50 & 100.86 \\
& Robust ($k=1.345$) & 102.70 & 103.07 & 100.42 \\
& Robust ($k=1.2$) & 102.35 & 102.21 & 100.84 \\
\midrule
\multirow{6}{*}{$0.3$}
& OS & 100.00 & 100.00 & 100.00 \\
& UN & 98.64 & 100.95 & 101.05 \\
& Robust ($k=1.5$) & 104.41 & 103.33 & 101.23 \\
& Robust ($k=1.345$) & 103.75 & 103.09 & 101.05 \\
& Robust ($k=1.2$) & 103.62 & 102.68 & 101.55 \\
\bottomrule
\end{tabular}
\end{table}

\begin{table}[htbp]
\centering
\footnotesize
\caption{Relative efficiency ratios (ER) of each method compared with OS under the alternative hypothesis ($\lambda_1=1.5$, $\lambda_2=2$, $\phi_1=0.6$, $\phi_2=2$) for the one-sided test ($\alpha=0.025$).}
\label{tab:er_vs_os_nb_unequal}
\begin{tabular}{llccc}
\toprule
& & \multicolumn{3}{c}{$n_1$} \\
\cmidrule(lr){3-5}
$\epsilon$ & Method & 5 & 10 & 50 \\
\midrule
\multirow{6}{*}{$0$}
& OS & 100.00 & 100.00 & 100.00 \\
& UN & 98.70 & 101.43 & 101.10 \\
& Robust ($k=1.5$) & 103.42 & 103.02 & 103.74 \\
& Robust ($k=1.345$) & 104.50 & 104.14 & 103.87 \\
& Robust ($k=1.2$) & 104.69 & 104.34 & 104.33 \\
\midrule
\multirow{6}{*}{$0.1$}
& OS & 100.00 & 100.00 & 100.00 \\
& UN & 98.84 & 101.44 & 101.15 \\
& Robust ($k=1.5$) & 103.36 & 102.33 & 102.66 \\
& Robust ($k=1.345$) & 103.46 & 102.87 & 102.88 \\
& Robust ($k=1.2$) & 104.70 & 103.49 & 102.69 \\
\midrule
\multirow{6}{*}{$0.2$}
& OS & 100.00 & 100.00 & 100.00 \\
& UN & 99.67 & 100.48 & 101.15 \\
& Robust ($k=1.5$) & 103.36 & 102.13 & 101.96 \\
& Robust ($k=1.345$) & 103.64 & 102.69 & 101.90 \\
& Robust ($k=1.2$) & 104.74 & 102.92 & 101.85 \\
\midrule
\multirow{6}{*}{$0.3$}
& OS & 100.00 & 100.00 & 100.00 \\
& UN & 98.90 & 100.98 & 100.77 \\
& Robust ($k=1.5$) & 103.67 & 102.41 & 102.07 \\
& Robust ($k=1.345$) & 103.50 & 101.74 & 101.99 \\
& Robust ($k=1.2$) & 104.86 & 102.97 & 102.08 \\
\bottomrule
\end{tabular}
\end{table}

\section{Sample Size Planning Formula}

Tables~\ref{table-s1} and~\ref{table-s2} present the sample size planning formulas and the corresponding blinded sample size re-estimation settings for the examples described above.

\begin{table}[htbp]
\centering
\caption{Sample Size Planning Formula and Blinded Sample Size Re-estimation}
\label{table-s1}
\resizebox{\textwidth}{!}{
\begin{tabular}{p{3.5cm} p{4.5cm} p{10.5cm}}
\toprule
Distribution & Model specification & Sample size planning formula \\
\midrule
Normal
&
\makecell[l]{
$X_i \sim \mathcal{N}(\mu_1,\sigma^2)$ \\
$Y_j \sim \mathcal{N}(\mu_2,\sigma^2)$
}
&
Test $H_0:\mu_1-\mu_2=0$ vs.\ $H_1:\mu_1-\mu_2=\delta_1$.
\[
n=\frac{(z_{1-\alpha}+z_{1-\beta})^2}{\delta_1^2}
\left(\frac{\sigma^2}{r}+\sigma^2\right)
\]
Blinded setting:
\begin{align*}
  \mu_1
    = g_1(\bar{\mu},\delta_1)
    = \bar{\mu} + \frac{\delta_1}{r+1},
  \
  \mu_2
    = g_2(\bar{\mu},\delta_1)
    = \bar{\mu} - \frac{r\,\delta_1}{r+1}.
\end{align*}
\\
\midrule
Binary
&
\makecell[l]{
$X_i \sim \mathrm{Bernoulli}(p_1)$\\
$Y_j \sim \mathrm{Bernoulli}(p_2)$
}
&
Test $H_0:p_1-p_2=0$ vs.\ $H_1:p_1-p_2=\delta_1$.
\[
n=\frac{(z_{1-\alpha}+z_{1-\beta})^2}{\delta_1^2}
\left(\frac{p_1(1-p_1)}{r}+p_2(1-p_2)\right)
\]
Blinded setting:
\begin{align*}
  p_1
    = g_1(\bar{p},\delta_1)
    = \bar{p} + \frac{\delta_1}{r+1},
  \
  p_2
    = g_2(\bar{p},\delta_1)
    = \bar{p} - \frac{r\,\delta_1}{r+1}.
\end{align*}
\\
\midrule
Negative binomial
&
\makecell[l]{
$X_i \sim \mathrm{NB}(\lambda_1 t_{1i},\phi)$\\
$Y_j \sim \mathrm{NB}(\lambda_2 t_{2j},\phi)$
}
&
Test $H_0:\log\lambda_1-\log\lambda_2=0$
vs.\ $H_1:\log\lambda_1-\log\lambda_2=\delta_1$.
\[
n=\frac{(z_{1-\alpha}+z_{1-\beta})^2}{\delta_1^2}
\left(\frac{\tau_1^2}{r}+\tau_2^2\right)
\]
where
\[
\tau_1^2=
\frac{
\left(\frac{1}{m_1}\sum_{i=1}^{m_1} t_{1i}\right)\lambda_1^{-1}
+\left(\frac{1}{m_1}\sum_{i=1}^{m_1} t_{1i}^2\right)\phi
}{
\left(\frac{1}{m_1}\sum_{i=1}^{m_1} t_{1i}\right)^2
},
\]
and
\[
\tau_2^2=
\frac{
\left(\frac{1}{n_1}\sum_{j=1}^{n_1} t_{2j}\right)\lambda_2^{-1}
+\left(\frac{1}{n_1}\sum_{j=1}^{n_1} t_{2j}^2\right)\phi
}{
\left(\frac{1}{n_1}\sum_{j=1}^{n_1} t_{2j}\right)^2
}.
\]
Blinded setting:
\begin{align*}
  \lambda_1
    = g_1(\bar{\lambda},\delta_1)
    = \frac{(r+1)\bar{\lambda}e^{\delta_1}}{re^{\delta_1}+1} ,
  \
  \lambda_2
    = g_2(\bar{\lambda},\delta_1)
    = \frac{(r+1)\bar{\lambda}}{re^{\delta_1}+1} .
\end{align*}
\\
\bottomrule
\end{tabular}
}
\end{table}

\begin{table}[htbp]
\centering
\caption{Sample Size Planning Formula and Blinded Sample Size Re-estimation}
\label{table-s2}
\resizebox{\textwidth}{!}{
\begin{tabular}{p{3.5cm} p{4.5cm} p{10.5cm}}
\toprule
Distribution & Model specification & Sample size planning formula \\
\midrule
Gamma
&
\makecell[l]{
$X_i \sim \mathrm{Gamma}(\mu_1,\phi)$\\
$Y_j \sim \mathrm{Gamma}(\mu_2,\phi)$
}
&
Test $H_0:\log\mu_1-\log\mu_2=0$
vs.\ $H_1:\log\mu_1-\log\mu_2=\delta_1$.
\[
n=\frac{(z_{1-\alpha}+z_{1-\beta})^2}{\delta_1^2}
\left(\frac{\phi}{r}+\phi\right)
\]
Blinded setting:
\begin{align*}
  \mu_1
    = g_1(\bar{\mu},\delta_1)
    = \frac{(r+1)\bar{\mu}}{re^{\delta_1}+1}e^{\delta_1},
    \
  \mu_2
    = g_2(\bar{\mu},\delta_1)
    = \frac{(r+1)\bar{\mu}}{re^{\delta_1}+1}.
\end{align*}
\\
\midrule
Inverse Gaussian
&
\makecell[l]{
$X_i \sim \mathrm{IG}(\mu_1,\lambda)$\\
$Y_j \sim \mathrm{IG}(\mu_2,\lambda)$
}
&
Test $H_0:\log\mu_1-\log\mu_2=0$
vs.\ $H_1:\log\mu_1-\log\mu_2=\delta_1$.
\[
n=\frac{(z_{1-\alpha}+z_{1-\beta})^2}{\delta_1^2}
\left(\frac{\mu_1}{r\lambda}+\frac{\mu_2}{\lambda}\right)
\]
Blinded setting:
\begin{align*}
  \mu_1
    = g_1(\bar{\mu},\delta_1)
    = \frac{(r+1)\bar{\mu}}{re^{\delta_1}+1}e^{\delta_1},
    \
  \mu_2
    = g_2(\bar{\mu},\delta_1)
    = \frac{(r+1)\bar{\mu}}{re^{\delta_1}+1}.
\end{align*}
\\
\midrule
Lognormal
&
\makecell[l]{
$X_i \sim \mathrm{Lognormal}(\mu_1,\sigma^2)$\\
$Y_j \sim \mathrm{Lognormal}(\mu_2,\sigma^2)$
}
&
Test $H_0:\mu_1-\mu_2=0$ vs.\ $H_1:\mu_1-\mu_2=\delta_1$.
\[
n=\frac{(z_{1-\alpha}+z_{1-\beta})^2}{\delta_1^2}
\left(\frac{\sigma^2}{r}+\sigma^2\right)
\]
Blinded setting:
\begin{align*}
  \mu_1
    = g_1(\bar{\mu},\delta_1)
    = \log\frac{(r+1)e^{\bar{\mu}}}{r e^{\delta_1} +1} + \delta_1,
    \
  \mu_2
    = g_2(\bar{\mu},\delta_1)
    = \log\frac{(r+1)e^{\bar{\mu}}}{r e^{\delta_1} +1}.
\end{align*}
\\
\bottomrule
\end{tabular}
}
\end{table}

\end{document}